\documentclass[onecolumn]{autart}    
\usepackage{amsmath} 
\usepackage{chngcntr}
\usepackage{amssymb}  
\usepackage{graphicx} 
\usepackage{mathptmx} 
\usepackage{times} 
\usepackage{datetime}
\usepackage{framed} 
\usepackage{comment}
\usepackage{enumerate}
\usepackage{color}
\usepackage{comment}
\usepackage{tikz}
\usepackage{arydshln} 
\usepackage{subdepth}
\def\<{\leqslant}           
\def\>{\geqslant}           

\def\d{\partial}
\def\wh{\widehat}
\def\wt{\widetilde}
\def\~{\wt{~}}
\def\Re{\mathrm{Re}}   
\def\Im{\mathrm{Im}}   
\def\O{\mathrm{O}}    
\def\Sp{\mathrm{Sp}}   

\def\cH{\mathcal{H}}     
\def\mA{\mathbb{A}}      
\def\mR{\mathbb{R}}      
\def\mC{\mathbb{C}}      
\def\rT{\mathrm{T}}        

\def\[[[{[\![\![}
\def\]]]{]\!]\!]}

\def\bra{{\langle}}
\def\ket{{\rangle}}

\def\rd{{\rm d}}        

\def\cL{\mathcal{L}}

\def\bA{\mathbf{A}}

\def\x{\times}
\def\ox{\otimes}

\def\fA{\mathfrak{A}}

\def\vec{\mathrm{vec}}

\def\mbT{\mathbfit{T}}

\def\cF{\mathcal{F}}

\def\cL{\mathcal{L}}

\def\cD{\mathcal{D}}

\def\cR{\mathcal{R}}
\def\cRH{\mathcal{RH}}

\def\cI{{\mathcal I}}
\def\cP{{\mathcal P}}
\def\cQ{{\mathcal Q}}

\def\cE{{\mathcal E}}

\def\cS{{\mathcal S}}

\def\sJ{\mathsf{J}}

\def\mH{\mathbb{H}}
\def\mS{\mathbb{S}}

\def\Ups{\Upsilon}

\def\diag{\mathop{\mathrm{diag}}}    

\def\argmin{\mathop{\mathrm{arg\, min}}}    

\DeclareMathAlphabet{\mathbfit}{OML}{cmm}{b}{it}
\DeclareMathAlphabet      {\mathbfit}{OML}{cmm}{b}{it}
\DeclareMathAlphabet      {\mathbfd}{OT1}{cmr}{bx}{n}
\DeclareMathAlphabet{\bit}{OML}{cmm}{b}{it}

\begin{document}
\begin{frontmatter}

\title{Quantum Linear Coherent Controller Synthesis: A Linear Fractional Representation Approach\thanksref{footnoteinfo}} 
\thanks[footnoteinfo]{Corresponding author: Arash~Kh.~Sichani.}
\author[UNSW]{Arash~Kh.~Sichani}\ead{arash\_kho@hotmail.com},\qquad    
\author[UNSW]{Ian~R.~Petersen}\ead{i.r.petersen@gmail.com}  
\address[UNSW]{School of Engineering and Information Technology, University of New South Wales, Canberra,
 ACT 2600, Australia}  

\begin{abstract}                          
This paper is concerned with a linear fractional representation approach to the synthesis of linear coherent quantum controllers for a given linear quantum plant. The plant and controller represent open quantum harmonic oscillators and are modelled by linear quantum stochastic differential equations. The feedback interconnections between the plant and the controller are assumed to be established through quantum bosonic fields. In this framework, conditions for the stabilization of a given linear quantum plant via linear coherent quantum feedback are addressed using a stable factorization approach. The class of all stabilizing quantum controllers is parameterized in the frequency domain. Coherent quantum weighted $\cH_2$ and $\cH_\infty$ control problems for linear quantum systems are formulated in the frequency domain. Finally, a projected gradient descent scheme is outlined for the coherent quantum weighted $\cH_2$ control problem.
\end{abstract}
\end{frontmatter}

\section{Introduction}
The main motivation for coherent quantum feedback control is based on avoiding the loss of quantum information in conversion to classical signals which occurs during measurement \cite{M_1998}\cite{L_2000}. This approach builds on the technique of constructing a feedback network from the interconnection of quantum systems, for example, through field coupling; see \cite{GJ09,GJN10}. In this framework, coherent quantum control theory aims at developing systematic methods to design measurement-free interconnections of Markovian quantum systems modelled by quantum stochastic differential equations (QSDEs); for example, see \cite{JNP_2008,NJP_2009,P_2010}.                       
Owing to recent advances in quantum optics, the implementation of quantum feedback networks governed by linear QSDEs \cite{P_1992,M_2008,P_2010} is possible using quantum-optical components, such as optical cavities, beam splitters and phase shifters, provided the former represent open quantum harmonic oscillators (OQHOs) with a quadratic Hamiltonian and linear system-field coupling operators with respect to the state variables satisfying canonical commutation relations \cite{EB_2005,GZ_2004}. This important class of linear QSDEs models the Heisenberg evolution of pairs of conjugate operators in a multi-mode quantum harmonic oscillator that is coupled to external bosonic fields. As a consequence, the notion of physical realizability (PR) addresses conditions under which a linear QSDE represents an OQHO. This condition is organised as a set of constraints on the coefficients of the QSDE \cite{JNP_2008} or, alternatively, on the quantum system transfer matrix \cite{SP_2012,AKhP16} in the frequency domain. These constraints complicate the solution of the coherent quantum synthesis problems which are otherwise reducible to tractable unconstrained counterparts in classical control theory.

    Coherent quantum feedback control problems, such as internal stabilization and optimal control design, are of particular interest in linear quantum control theory \cite{JNP_2008,P_2010}. These problems are amenable to transfer matrix design methods \cite{yan2003I,yan2003II,GJN10,P_2010,SP_2012}. Among the transfer matrix approaches to the control problems for linear multivariable systems, the linear fractional representation approach to analysis and synthesis has been largely developed in the literature; see \cite{vid2011} and the references therein.         
The linear fractional representation approach is a cornerstone in the study of stabilization problems. The central idea of this approach is to represent the transfer matrix of a plant as fractions of stable rational matrices to generate stable factorizations. By combining the idea of the stable factorizations of a plant with the concept of coprimeness, necessary and sufficient conditions for internal stabilizability are derived in terms of B\'{e}zout equations \cite{vid2011}. By solving these B\'{e}zout equalities, a parameterization of all stabilizing controllers, known as the Youla-Ku\v{c}era parameterization, is obtained. This idea gives rise to a method which leads to the solution of several important control problems; see for example \cite{vid2011}.

    The Youla-Ku\v{c}era parameterization was developed originally in the frequency domain for finite-dimensional linear time-invariant systems using transfer function methods, see \cite{youla76I,youla76II}, and generalized to infinite-dimensional systems afterwards \cite{desoer80,quadrat2003,vid2011}. The state space representation of all stabilizing controllers has also been addressed for finite-dimensional, linear time-invariant \cite{nett84} and time-varying \cite{dale93} systems, and the approach was shown to be applicable to a class of nonlinear systems \cite{hammeri85,paice90,anderson98}. In the Youla-Ku\v{c}era parameterization, the feedback loop involving the controller is redefined in terms of another parameter known as the Youla or $Q$ parameter. The closed-loop map is then an affine function of $Q$, and so the optimal $Q$ in standard optimal stabilization problems can be easily found. Moreover, some constraints, such as internal stability, are reduced to convex constraints on $Q$. Therefore, this approach provides a tool that allows us to better understand the dichotomy between tractable and intractable control synthesis problems in the presence of additional constraints on the controller; see for example \cite{Boyd91}.
     
    In the present paper, we employ a stable factorization approach in order to develop a counterpart of the classical Youla-Ku\v{c}era parameterization for describing the set of linear coherent quantum controllers that stabilize a linear quantum stochastic system (LQSS). In particular, we address the problem of coherent quantum stabilizability of a given linear quantum plant. The class of all stabilizing controllers is parameterized in the frequency domain. This approach allows weighted $\cH_2$ and $\cH_\infty$ coherent quantum control problems to be formulated for linear quantum systems in the frequency domain. In this way, the weighted $\cH_2$ and $\cH_\infty$ control problems are reduced to constrained optimization problems with respect to the Youla-Ku\v{c}era parameter with convex cost functionals. Moreover, these problems are organised as a constrained version of the model matching problem \cite{francis87}. Finally, a projected gradient descent scheme is proposed to provide a conceptual solution to the weighted $\cH_2$ coherent quantum control problem in the frequency domain.
    
    The rest of this paper is organised as follows. Section~\ref{sec:not} outlines the notation used in the paper. Linear quantum stochastic systems are described in Section~\ref{sec:QSS}. The coherent quantum feedback interconnection under consideration is described in Section~\ref{sec:intcnct}. Section~\ref{sec:LQHO_PR} revisits the PR conditions for linear quantum systems in the frequency domain. Sections~\ref{sec:cntlpara} and~\ref{sec:QYK} formulate a quantum version of the Youla-Ku\v{c}era parameterization and provide relevant preparatory material. Also, a class of unstabilizable LQSS systems is presented. Coherent quantum weighted $\mathcal{H}_2$ and $\mathcal{H}_\infty$ control problems are introduced in Section~\ref{sec:H2}. A projected gradient descent scheme for the quantum weighted $\mathcal{H}_2$ control problem is outlined in Section~\ref{sec:PGS}. Section~\ref{sec:Conclusion} gives concluding remarks. Appendix~\ref{app:sec:PR_LFT_INV} provides a parameterization of linear coherent quantum feedback systems in the position-momentum form. Appendix~\ref{App:CLD} provides a Cholesky-like factorization for skew-symmetric matrices. Appendices~ \ref{app:Appndx_LFT} and \ref{app:GBI} provide relevant facts about linear fractional transformations and the general B\'{e}zout identity. Appendices~\ref{app:JJYoulaSS} and~\ref{app:proj} provide complementary materials for purposes of Section~\ref{sec:QYK} and Section~\ref{sec:PGS}.  

A preliminary version of this work has been published in the conference proceedings of the 10th Asian control conference in 2015 \cite{AKh_2015}.  In comparison to the conference version, use is made of a modified version of the physical realizability condition for linear quantum stochastic systems in the frequency domain \cite{AKhP16} which leads to more complete and simple results. The changes include a real-valued parameterization of the linear coherent quantum stochastic feedback systems (without loss of generality) and the omission of technical assumptions in the main results of the paper. The main theorem, Theorem~8, in \cite{AKh_2015} and its proof has been modified to provide a parameterization of the set of \emph{all} stabilizing linear coherent quantum controllers. A class of linear quantum systems is presented which cannot be stabilized by linear coherent quantum controllers. Complementary results and technical details are presented in the appendices.

\section{Notation}\label{sec:not}
Unless specified otherwise,  vectors are organized as columns, and the transpose $(\cdot)^{\rT}$ acts on matrices with operator-valued entries as if the latter were scalars. For a vector $X$ of self-adjoint operators $X_1, \ldots, X_r$ and a vector $Y$ of operators $Y_1, \ldots, Y_s$, the commutator matrix is defined as an $(r\x s)$-matrix
$
    [X,Y^{\rT}]
    :=
    XY^{\rT} - (YX^{\rT})^{\rT}
$
whose $(j,k)$th entry is the commutator
$
    [X_j,Y_k]
    :=
    X_jY_k - Y_kX_j
$ of the operators $X_j$ and $Y_k$. Furthermore, $(\cdot)^{\dagger}:= ((\cdot)^{\#})^{\rT}$ denotes the transpose of the entry-wise operator adjoint $(\cdot)^{\#}$. When it is applied to complex matrices,  $(\cdot)^{\dagger}$ reduces to the complex conjugate transpose  $(\cdot)^*:= (\overline{(\cdot)})^{\rT}$. 
The positive semi-definiteness of matrices is denoted by $\succcurlyeq$, and $\ox$ is the tensor product of spaces or operators (for example, the Kronecker product of matrices). Furthermore, $\mS_r$, $\mA_r$
 and
$
    \mH_r
    :=
    \mS_r + i \mA_r
$ denote
the subspaces of real symmetric, real antisymmetric and complex Hermitian  matrices of order $r$, respectively, with $i:= \sqrt{-1}$ the imaginary unit. Also, $I_r$ denotes the identity matrix of order $r$, the identity operator on a space $\cH$ is denoted by $\cI_{\cH}$, and the matrices 
$\sJ:={\scriptsize
	\begin{bmatrix}
		 0 & 1\\
		-1 & 0
	\end{bmatrix}}$
and
$J_{r}:=I_{\frac{r}{2}} \ox \sJ$. 
The sets $\O(r) := \big\{\Sigma \in \mR^{r\x r}: \Sigma^\rT \Sigma =I \big\}$ and $\Sp(r,\mR) := \big\{\Sigma \in \mR^{r\x r}: \Sigma^\rT J_{r} \Sigma =J_{r} \big\}$ refer to the group of orthogonal matrices and the group of symplectic real matrices of order $r$. 
The notation
$
    {\scriptsize\left[
    \begin{array}{c|c}
          A & B \\
          \hline
          C & D
    \end{array}
    \right]}
$
refers to a state-space realization of the corresponding transfer matrix $\Gamma(s) := C(sI-A)^{-1}B+D$ with a complex variable $s \in \mC$.
The conjugate system transfer function $(\Gamma(-\overline{s}))^*$ is written as $\Gamma^{\sim}(s)$. 
The Hardy space of (rational) transfer functions of type $p=2,\infty$ is denoted by $\cH_p$ (respectively, $\cRH_p$). The symbol $\ox$ is used for the tensor product of spaces. 
\section{Linear Quantum Stochastic Systems}\label{sec:QSS}
	We consider a Markovian quantum stochastic system interacting with an external boson field. The system has $n$ dynamic variables $X_{1}(t), \ldots, X_{n}(t)$, where $t\>0$ denotes time. We generally suppress the time argument of operators, unless we are explicitly concerned with their time dependence, with the understanding that all operators are evaluated at the same time. The system variables are self-adjoint operators on an underlying complex separable Hilbert space $\cH$ which satisfy the Heisenberg canonical commutation relations (CCRs)
\begin{equation}
\label{xCCR_1}
    [X, X^{\rT}] = 2i \Theta \otimes \cI_{\cH}, 
    \qquad 
    X:=
    \begin{bmatrix}
        X_{1}\\
        \vdots\\
        X_{n}
    \end{bmatrix},
\end{equation}
on a dense domain in $\cH$, where $\theta \in \mA_n$ is nonsingular. In what follows, the matrix $\Theta \otimes \cI_{\cH}$ will be identified with $\Theta$.
The system variables evolve in time according to a Hudson-Parthasarathy QSDE \cite{P_1992} with identity scattering matrix (which eliminates from consideration the gauge, also known as conservation, processes associated with photon exchange between the fields):
\begin{equation}
	\label{dx_f_g}
	\rd X = f \rd t + g \rd W.
\end{equation}
The $n$-dimensional drift vector $f$ and the dispersion $(n \times m)$-matrix $g$ of the QSDE (\ref{dx_f_g}) are given by
\begin{equation}
	\label{drift_dis}
	f:= \cL(X)=i[H,X]+\cD(X), 
	\quad 
	g:= -i [ X, L^\rT], 
	\quad 
	L:= 
	\begin{bmatrix}
		L_1 \\ \vdots \\ L_{m}
	\end{bmatrix}.
\end{equation}
Here, $H$ is the system Hamiltonian which is usually represented as a function of the system variables, and $L_1, \hdots, L_{m}$ are the system-field coupling operators (the dimension $m$ is assumed to be even). These self-adjoint operators act on the space $\cH$ and specify the self-energy of the system and its interaction with the environment. Furthermore, $\cL$ coincides with the Gorini-Kossakowski-Sudarshan-Lindblad (GKSL) generator \cite{GKS_1976,L_1976} which acts on a system operator $\xi$ as 
\begin{equation}
	\label{GKSL}
	\cL(\xi):= i[H, \xi]+ \cD(\xi)
\end{equation}
and is evaluated entry-wise at the vector $X$ in (\ref{xCCR_1}), and $\cD$ is the decoherence superoperator given by
\begin{align}	
	\nonumber
	\cD(\xi)
	:=&
	\frac{1}{2} \sum_{j,k=1}^{m} \omega_{jk}\big([L_j,\xi]L_k+L_j[\xi,L_k]\big)\\ 
	\label{decoh}
	=&\frac{1}{2}\big([L^\rT,\xi]\Omega L + L^\rT \Omega [\xi,L]\big).
\end{align}
In the QSDE (\ref{dx_f_g}), $W$ is an $m$-dimensional vector of quantum Wiener processes $W_{1}, \ldots, W_{m}$, which are self-adjoint operators on a boson Fock space \cite{H_1991,P_1992}, modelling the external fields with a  positive semi-definite It\^{o} matrix $\Omega:=\big( \omega_{jk} \big)_{1\<j,k\<m} \in \mH_{m}$:
\begin{equation}
\label{WW_1}
    \rd W \rd W^{\rT}
    =
    \Omega \rd t.
\end{equation}  
The entries of $W$ are linear combinations of the field annihilation $\fA_1, \hdots, \fA_{\frac{m}{2}}$ and creation $\fA_1^\dagger, \hdots, \fA_{\frac{m}{2}}^\dagger$ operator processes \cite{HP_1984,P_1992}:
\begin{equation}
	\label{QWP}
	W:=2
	P_m	
	\begin{bmatrix}
		\Re \fA\\
		\Im \fA
	\end{bmatrix}		
	=
	P_m
	T_m
	\begin{bmatrix}
		\fA\\
		\fA^\#
	\end{bmatrix}
	,
	\quad	
	T_m:=
	\begin{bmatrix}
		1  & 1\\
		-i & i
	\end{bmatrix}	
	\ox
	I_{\frac{m}{2}}	
	,	
\end{equation}
where 	
$$
	\fA
	:=
	\begin{bmatrix}
		\fA_1\\
		\vdots\\
		\fA_\frac{m}{2}
	\end{bmatrix}
	,
	\quad 
		\fA^\#
	:=
	\begin{bmatrix}
		\fA_1^\dagger\\
		\vdots\\
		\fA_\frac{m}{2}^\dagger
	\end{bmatrix},	
$$	
with the quantum It\^{o} relations
\begin{equation*}
	\rd 
	\begin{bmatrix}
		\fA\\
		\fA^\#
	\end{bmatrix}
	\rd
	\begin{bmatrix}
		\fA^\dagger 
		&
		\fA^\rT
	\end{bmatrix}	
	:=
	\begin{bmatrix}
		\rd \fA \rd \fA^\dagger	   &	\rd \fA \rd \fA^\rT\\
		\rd \fA^\# \rd \fA^\dagger &    \rd \fA^\# \rd \fA^\rT
	\end{bmatrix}
	=
	\Bigg(
	\begin{bmatrix}
		1 & 0\\
		0 & 0
	\end{bmatrix}
	\ox I_{\frac{m}{2}}
	\Bigg)
	\rd t,
\end{equation*}
where $P_m \in \mR^{m \times m}$ is a permutation matrix such that
$P_m(\sJ \ox I_{\frac{m}{2}})P_m^\rT=I_{\frac{m}{2}} \ox \sJ=J_m$.
Accordingly, the It\^{o} matrix $\Omega$ in (\ref{WW_1}) is described by
\begin{equation}
	\Omega=
	P_m
	T_m
	\bigg({\scriptsize
	\begin{bmatrix}
		1 & 0\\
		0 & 0
	\end{bmatrix}}
	\ox I_{\frac{m}{2}}
	\bigg)
	T_m^*
	P_m^\rT
	=
	I_{m}+iJ_{m}
	=
    \Omega^*\succcurlyeq 0
	.
\end{equation}
In accordance with the evolution (\ref{dx_f_g}), the system variables $X_1(t),\allowbreak \hdots,\allowbreak X_n(t)$ at any given time $t \>0$ act on a tensor product Hilbert space $\cH_0 \otimes \cF_t$, where $\cH_0$ is the initial complex separable Hilbert space of the system and $\cF_t$ is the Fock filtration. By using the quantum stochastic calculus \cite{P_1992}, in accordance with (\ref{drift_dis})--(\ref{decoh}), the QSDE (\ref{dx_f_g}) is derived from the Heisenberg unitary evolution on the tensor product of system and field spaces $\cH=\cH_0 \otimes \cF_t$ described by the quantum stochastic flow
\begin{equation}
	\label{X_t}
	X(t)=U(t)^\dagger ( X(0) \otimes \cI_\cF) U(t),
\end{equation}
where the unitary operator $U(t)$ satisfies the initial condition $U(0)=\cI_\cH$ and is governed by a stochastic Schr\"{o}dinger equation 
\begin{equation}
	\label{U_Sch}
	\rd U(t)=
	-
	\Big( 
		\big( 
		iH(0) + \frac{1}{2} L(0)^{\rT} \Omega L(0) 
		\big) 
		\rd t 
		+  
		i L(0)^{\rT} \rd W(t) 
	\Big) 
	U(t).
\end{equation}
The unitary evolution in (\ref{U_Sch}) preserves the CCRs (\ref{xCCR_1}): 
\begin{align*}
	[X(t),X(t)^\rT]
	&=
	U(t)^\dagger([X(0),X(0)^\rT] \otimes \cI_\cF)U(t)
	\\
	&=
	2i\Theta U(t)^\dagger \cI_{\cH_0 \otimes \cF} U(t) 
	= 
	2i \Theta,
\end{align*}
where the entries of $X(0)$ commute with those of $W(t)$ as operators on different spaces.

In particular, by considering the following Hamiltonian and system-field coupling operators 
\begin{equation}
	\label{H_M}
    H = \frac{1}{2} X^{\rT} R X = \frac{1}{2}\sum_{j,k=1}^{n}r_{jk}X_jX_k, 
    \qquad L=MX,
\end{equation}
the system corresponds to an $\frac{n}{2}$-mode OQHO \cite{EB_2005,GZ_2004}. Here, $R:= (r_{jk})_{1\< j,k\< n}$ is a real symmetric matrix of order $n$, and $M\in \mR^{m\times n}$ is the system-field coupling parameter. By substituting the Hamiltonian and coupling operators from (\ref{H_M}) into (\ref{drift_dis}) and using the CCRs (\ref{xCCR_1}), it follows that the QSDE takes the form
\begin{align}
\label{equ:tdomain_model:1}
    \rd
    X(t)
    &= A X(t) \rd t+
       B \rd W(t),
\end{align}\noindent
where, in view of (\ref{dx_f_g}), the drift vector $f=AX$ and the dispersion matrix $g=B$ are given by
\begin{equation}
	\label{A_B}
	A:=2\Theta R - \frac{1}{2} BJ_{m}B^{\rT} \Theta^{-1},\qquad B:= 2 \Theta M^\rT.
\end{equation} 
The term $-\frac{1}{2} BJ_m B^{\rT} \Theta^{-1} X$ in the drift represents the GKSL decoherence superoperator which acts on the system variables and is associated with the system-field interaction.

\begin{figure}[htpb]
\begin{center}
\unitlength=1mm
\linethickness{0.6pt}
\hskip-10mm
\begin{picture}(0,25)
    \put(0,15){\framebox(10,10)[cc]{\scriptsize$S$}}
    \put(-8,20){\vector(1,0){8}}
    \put(10,20){\vector(1,0){8}}
    \put(-9,20){\makebox(0,0)[rc]{{\scriptsize$W$}}}
    \put(22,20){\makebox(0,0)[rc]{{\scriptsize$Y$}}}
\end{picture} \vskip-15mm
\caption[Input-output block-diagram of open quantum systems.]{Input-output block-diagram.}
\label{fig:in_out}
\end{center}
\end{figure}
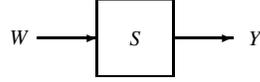

We associate a vector $Y$ of output field dynamic variables $Y_1, \hdots, Y_p$ with an OQHO:
\begin{equation}
	\label{equ:tdomain_model:2}
	\rd Y = C X \rd t + D \rd W,
\end{equation}
where $C \in \mR^{p \times n}$ with the dimension $p$ assumed to be even and the full row rank matrix $D \in \mR^{p \times m}$ satisfies 
\begin{equation}
	\label{rD}
	D(I_{m}+iJ_{m}) D^\rT = I_{p}+iJ_{p}
\end{equation}
from a similar condition to (\ref{WW_1}) for the outputs ($p \< m$). The output field satisfies the non-demolition condition \cite{B_1989} with respect to the dynamic variables in the sense that
\begin{equation}
	\label{Nondemc}
	[X(t),Y(s)^{\rT}]=0,
    \qquad
    t\> s.
\end{equation}
Due to this non-demolition property, the output fields can be interpreted as ideal observations of the open quantum system, except that $Y_1, \hdots, Y_p$ do not commute with each other. This condition implies an algebraic relation between $C$ and $D$ in (\ref{equ:tdomain_model:2}):
\begin{align}
	\label{rCDB}
	C =  -D J_{m} B^\rT \Theta^{-1}.
\end{align}
We will refer to the input-output dynamics of a system $S$ which is described by (\ref{equ:tdomain_model:1}) and (\ref{equ:tdomain_model:2}) as a LQSS. This system can be parameterized by the triple $(D,M,R)$ in (\ref{A_B}) and (\ref{rCDB}). By analogy with classical linear systems, we often represent this system by the state-space realization
$
S=
\scriptsize
\left[
	\begin{array}{c c}
		 A & B\\
		 C & D\\
	\end{array}	
\right]
$.
The input-output block-diagram of the LQSS is depicted as in Fig.~\ref{fig:in_out}.
Note that there exists a one-to-one correspondence between the real-valued $(D, M, R)$ parameterization of the LQSS, which will be referred to as the \emph{position-momentum form}, and the complex-valued, but structured, parameterization, referred to as \emph{annihilation-creation form} \cite{AKhP16}.

\section{Coherent Quantum Feedback Interconnection}
\label{sec:intcnct}
In this section, we will provide a framework for the interconnection of two LQSSs, one acting as the quantum plant and the other as the controller. In this framework, we consider field interconnections between the two systems with initial complex separable Hilbert spaces $\cH_1$, $\cH_2$. In particular, the vectors of dynamic variables of the plant and the controller are denoted by $X_1$ and $X_2$ (which consist of self-adjoint operators on the product space  $\cH_1\ox \cH_2\ox \cF_1\ox \cF_2$ at any subsequent moment of time $t>0$) and are assumed to satisfy CCRs
\begin{equation}
\label{CCRPK}
    [X,X^{\rT}]
    =
    2i\Theta,
    \qquad
    X
	:=
	\begin{bmatrix}
		X_1\\
		X_2
	\end{bmatrix}, 
    \qquad
    \Theta
    :=
    {\begin{bmatrix}
        \Theta_1 & 0 \\
        0 & \Theta_2
    \end{bmatrix}},
\end{equation}
where $\Theta_1, \Theta_2 \in \mA_{n}$ are constant nonsingular matrices.

\begin{figure}[htpb]
\begin{center}
\unitlength=1mm
\linethickness{0.6pt}
\begin{picture}(30.00,18.00)
    \put(10,10){\framebox(10,10)[cc]{\scriptsize$P$}}
    \put(2,12){\vector(1,0){8}}
    \put(2,18){\vector(1,0){8}}
    \put(20,12){\vector(1,0){8}}
    \put(20,18){\vector(1,0){8}}
    \put(0,12){\makebox(0,0)[rc]{{\scriptsize$W_u$}}}
    \put(30,12){\makebox(0,0)[lc]{{\scriptsize$Y_y$}}}
    \put(0,18){\makebox(0,0)[rc]{{\scriptsize$W_r$}}}
    \put(30,18){\makebox(0,0)[lc]{{\scriptsize$Y_z$}}}
\end{picture}\vskip-9mm
\caption[Partitioning of vectors of input and output field operators.]{This diagram depicts the way in which the input and output field conjugate pairs of the system $P$ are partitioned in (\ref{io_part}). This structure allows for field coupling to another quantum system.}\vskip-2mm
\label{fig:modifiedsystem}
\end{center}
\end{figure}
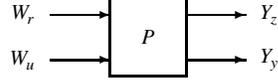

\begin{figure}[htpb]
\begin{center}
\unitlength=1mm
\linethickness{0.6pt}
\vskip.6cm
\begin{picture}(30.00,33.00)
    \put(10,25.5){\framebox(10,10)[cc]{\scriptsize$P$}}
    \put(10,10){\framebox(10,10)[cc]{\scriptsize$K$}}
    \put(2,33){\vector(1,0){8}}
    \put(20,33){\vector(1,0){8}}
    \put(2,28){\vector(1,0){8}}
    \put(20,28){\line(1,0){8}}
    \put(28,28){\line(0,-1){13}}
    \put(28,15){\vector(-1,0){8}}
    \put(10,15){\line(-1,0){8}}
    \put(2,15){\line(0,1){13}}
    \put(0,15){\makebox(0,0)[rc]{{\scriptsize$W_u$}}}
    \put(30,15){\makebox(0,0)[lc]{{\scriptsize$Y_y$}}}
    \put(0,33){\makebox(0,0)[rc]{{\scriptsize$W_r$}}}
    \put(30,33){\makebox(0,0)[lc]{{\scriptsize$Y_z$}}}
\end{picture}\vskip-8mm
\caption[A fully quantum closed-loop interconnection.]{This diagram depicts the fully quantum closed-loop system  which is the interconnection of the quantum systems $P$ and $K$. The effect of the environment on the closed-loop system is represented by $W_r$.}
\label{fig:system}
\end{center}
\end{figure}
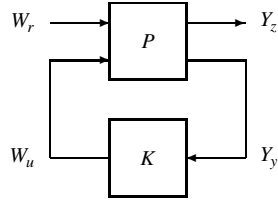

\begin{figure}[htpb]
\begin{center}
\unitlength=1mm
\linethickness{0.6pt}
\begin{picture}(30.00,39.00)
    \put(3.5,23){\dashbox(21,18)[cc]{}}
    \put(10,27){\framebox(10,10)[cc]{\scriptsize$P_1$}}
    \put(17.5,27){\makebox(10,10)[cc]{\scriptsize$P$}}    
    \put(10,10){\framebox(10,10)[cc]{\scriptsize$K$}}
    \put(5.5,35){\vector(1,0){4.5}}
    \put(-5,35){\line(1,0){9.5}}

    \put(-5,33){\line(1,0){9.5}}
    \put(5.5,33){\line(1,0){1}}    
    \put(6.5,33){\line(0,-1){8}}        
    \put(6.5,25){\line(1,0){15.5}}            
    \put(22,25){\line(0,1){2}}                
    \put(22,27){\line(1,0){6}}                    
    \put(28,27){\line(0,-1){11}}                        
    
    \put(20,35){\vector(1,0){12}}
    \put(7,29){\vector(1,0){3}}
    \put(20,29){\line(1,0){10}}
    \put(30,29){\line(0,-1){15}}
    \put(28,16){\vector(-1,0){8}}    
    \put(30,14){\vector(-1,0){10}}
    \put(10,14){\line(-1,0){10}}
    \put(10,16){\line(-1,0){8}}    
    \put(2,16){\line(0,1){13}}        
    \put(2,29){\line(1,0){4}}            
    
    \put(0,14){\line(0,1){17}}
    \put(0,31){\line(1,0){5}}    
    \put(5,31){\line(0,1){8}}        
    \put(5,39){\line(1,0){17}}            
    \put(22,39){\line(0,-1){2}}                
    \put(22,37){\vector(1,0){10}}                    
    
    \put(23,17){\makebox(0,0)[lb]{{\scriptsize$y_1$}}}
    \put(23,13){\makebox(0,0)[lt]{{\scriptsize$y_2$}}}    
    \put(3,17){\makebox(0,0)[lb]{{\scriptsize$u_1$}}}
    \put(3,13){\makebox(0,0)[lt]{{\scriptsize$u_2$}}}
    
    \put(-6,35){\makebox(0,0)[rb]{{\scriptsize$r_1$}}}
    \put(-6,33){\makebox(0,0)[rt]{{\scriptsize$r_2$}}}    
    \put(33,37){\makebox(0,0)[lb]{{\scriptsize$z_1$}}}    
    \put(33,35){\makebox(0,0)[lt]{{\scriptsize$z_2$}}}
\end{picture}\vskip-8mm
\caption[Conventional coherent quantum feedback interconnection.]{This diagram depicts the way in which the quantum system $P$, the concatenation of $P_1$ and the feedthroughs, is formed by grouping the exogenous inputs and outputs of the closed-loop system.}
\label{fig:routing}
\end{center}
\end{figure}
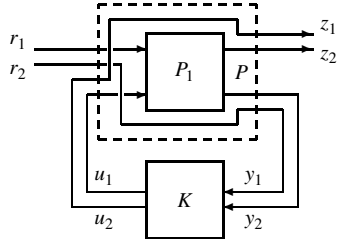

By analogy with similar structures in the interconnections arising in classical control theory, we partition the vectors $W$ and $Y$ of the plant $P$ input and output field operators in accordance with Fig.~\ref{fig:modifiedsystem}:
\begin{equation}
\label{io_part}
		W =
        {\scriptsize\begin{bmatrix}
			W_r\\
			W_u
		\end{bmatrix}},
    \qquad
		Y=
		{\scriptsize\begin{bmatrix}
			Y_z\\
			Y_y
		\end{bmatrix}}.
\end{equation}\noindent
Here $W_r$, $Y_z$, $W_u$, $Y_y$ denote the vectors of conjugate pairs of the input and output fields of the closed-loop system, and the input and output of $K$, which correspond to the classical reference, output, control and observation signals, respectively.
\vspace{3mm}
The field coupling feedback interconnection of the systems $P$ and $K$ is shown in Fig.~\ref{fig:system}.
Note that, similarly to the classical case, the interconnection in Fig.~\ref{fig:system} provides a general framework for the feedback interconnection of two quantum systems.
Note that in the case of quantum control, both the plant and controller will have exogenous inputs and outputs. However, this situation can be handled in the framework of Figure~\ref{fig:system}. Indeed, this framework includes the conventional coherent quantum feedback interconnection shown in Fig.~\ref{fig:routing}, where $P_1$ and $K$ act as the quantum plant and the controller, respectively. In this figure, the exogenous inputs and outputs of the closed-loop system are grouped together. Here, $r_1$ and $r_2$ represent the exogenous inputs of the plant and controller, respectively. Also, $z_1$ and $z_2$ represent the exogenous outputs of the controller and the plant, respectively.
\vskip5mm
Note that the feedback system in Fig.~\ref{fig:system} represents a LQSS; see Lemma~\ref{lem:PR_LFT_inv} in Appendix~\ref{app:sec:PR_LFT_INV}. 
\section[Open Quantum Harmonic Oscillators and Physical Realizability]{Open Quantum Harmonic Oscillators in the Frequency Domain and Physical Realizability}\label{sec:LQHO_PR} 


In accordance with Section~\ref{sec:QSS}, we consider the dynamics of the joint evolution of an $\frac{n}{2}$-mode OQHO and the external bosonic fields in the Heisenberg picture, represented by the linear QSDEs (\ref{equ:tdomain_model:1}) and (\ref{equ:tdomain_model:2}). Here, in view of the $(D,M,R)$ parameterization of LQSSs provided in Section~\ref{sec:QSS}, the matrices $A \in \mR^{n\x n}$, $B \in \mR^{n\x m}$, $C \in \mR^{m\x n}$, $D \in \mR^{m\x m}$ in (\ref{equ:tdomain_model:1}) and (\ref{equ:tdomain_model:2}) are given by
\begin{align}
    \label{equ:ABCD}
    {\scriptsize\begin{bmatrix}
        A & B\\
        C & D
    \end{bmatrix}}
    := 
    {\scriptsize\begin{bmatrix}
        2\Theta R - \frac{1}{2} BJ_{m}B^{\rT} \Theta^{-1} & B\\
        -D J_{m} B^\rT \Theta^{-1} & D
    \end{bmatrix}},
    \qquad
    B:=2 \Theta M^\rT,
%
\end{align}\noindent
where $\Theta \in \mA_{n}$ is the CCR matrix. Also, the parameter $R$ is a real symmetric matrix of order $n$ associated with the quadratic Hamiltonian of the OQHO, $M\in R^{m\times n}$ is the system-field coupling parameter, the feedthrough real matrix $D$ belongs to the subgroup of orthogonal symplectic matrices (the maximum compact subgroup of symplectic matrices) 
\begin{equation}
	\label{Spm}
	\Sp(m) = \O(m) \cap \Sp(m, \mR).
\end{equation}
The input-output map of the LQSS, governed by the linear QSDEs (\ref{equ:tdomain_model:1}) and (\ref{equ:tdomain_model:2}), is completely specified by a transfer function which is defined in the standard way as
    \begin{equation}
    \label{equ:hoc_freq}
    \Gamma(s) :=
    {\scriptsize\left[
    \begin{array}{c|c}
          A & B\\
          \hline
          C & D
    \end{array}
    \right]}=C(sI-A)^{-1}B+D,
    \end{equation}\noindent
where the matrices $A, B, C, D$ are parameterized by the triple $(D,M,R)$ as in (\ref{equ:ABCD}) with a given CCR matrix $\Theta$. As discussed above, in view of the specific structure of this parameterization, not every linear system, or system transfer function (\ref{equ:hoc_freq}) with an arbitrary quadruple $(A,B,C,D)$, represents the dynamics of a LQSS. This fact is addressed in the form of PR conditions for the quadruple $(A,B,C,D)$ to represent such an oscillator; see \cite{JNP_2008} for more details. The notion of PR for a transfer function is defined as follows.
\begin{defn} \cite{AKhP16} \label{def:PR_TF}
	The transfer function $\Gamma(s)$ is said to be \emph{physically realizable} if $\Gamma(s)$ represents a 
	LQSS, that is, there exists a minimal state-space realization for $\Gamma(s)$ which can be parameterized 
	by a triple $(D,M,R)$ as in (\ref{equ:ABCD}) for a given CCR matrix $\Theta$.
\end{defn}
Note that, in view of the results of Lemma~\ref{lem:ch_fact} in Appendix~\ref{App:CLD}, the invariance of transfer functions with respect to similarity transformations on the corresponding state-space realizations \cite{ZDG_1996} and Definition~\ref{def:PR_TF}, 
it can be shown that $\Gamma(s)$ is also physically realizable if there exists a minimal state-space realization for $\Gamma(s)$ which can be parameterized by the triple $(D,M,R)$ as in (\ref{equ:ABCD}) with any non-singular skew-symmetric matrix $\Theta$. The following lemma provides a PR condition for transfer matrices of linear quantum systems, which can be considered as a modified version of Theorem~4 in \cite{SP_2012}. The proof of this lemma is similar to the corresponding one in \cite[Theorem~1]{AKhP16} which is omitted for brevity. In what follows, the subscripts in $I_{m}$ and $J_{m}$ will often be omitted for brevity.
\begin{lem}
\label{lem:PR_Freq}
A transfer function $\Gamma(s)$ is physically realizable if and only if
	\begin{equation}
		\label{equ:JJUnit}
		\Gamma^{\sim}(s) J \Gamma(s)=J
	\end{equation}\noindent
for all $s\in \mC$, and the feedthrough matrix $D = \Gamma(\infty)$ is orthogonal.
\end{lem}
A transfer function $\Gamma (s)$, satisfying the condition (\ref{equ:JJUnit}), is said to be $(J,J)$-unitary; see, for example, \cite{SP_2012} and references therein. Since we consider this property for invertible square transfer matrices, in view of the fact that $J^2=-I$, the $(J,J)$-unitary condition (\ref{equ:JJUnit}) is equivalent to its dual form \cite{AKh_2015}:
\begin{equation}
		\label{dual}
		\Gamma(s) J \Gamma^{\sim}(s)=J.
\end{equation}	
	In view of the one-to-one correspondence described in \cite{AKhP16}, the results in Lemma~\ref{lem:PR_Freq} imply the results in \cite[Theorem~4]{SP_2012}. However, in comparison to \cite[Theorem~4]{SP_2012}, no additional technical assumptions are required in Lemma~\ref{lem:PR_Freq}. The technical assumption which is used in \cite{SP_2012} is referred to as spectral genericity of the linear quantum systems \cite{AKh_2015}.
\section{Parameterizations of All Stabilizing Controllers}\label{sec:cntlpara}
We consider a linear quantum plant and a linear quantum controller with square transfer matrices $P$ and $K$, respectively, each representing a LQSS in the frequency domain. In what follows, the argument $s$ of transfer functions will often be omitted for brevity. 

Following the field coupling feedback scheme introduced in Section~\ref{sec:intcnct}, we assume that the plant and the controller are connected according to Fig.~\ref{fig:system}. To allow for the feedback interconnection, we partition the plant transfer matrix as
\begin{align}
		\label{equ:modplnt}
        P       
        =        
        {\scriptsize \left [
        \begin{array} {c c}
            P_{11} & P_{12}\\
            P_{21}&P_{22}
        \end{array}
        \right ]} 
     	=:
	    {\scriptsize\left [
	    \begin{array}{c|c c}
	          A & B_1 & B_2\\
	          \hline
	          C_1 & D_{11} & D_{12} \\
	          C_2 & D_{21} & D_{22}
	    \end{array}
	    \right]}
		= 
		{\scriptsize \left [
        \begin{array} {c | c}
            A & B\\
            \hline
            C & D
        \end{array}
        \right ]} 	    
	    .
\end{align}
\noindent
The closed-loop transfer matrix between the exogenous inputs and outputs of interest can be calculated through the lower linear fractional transformation (LFT) of the plant and the controller in the frequency domain \cite{ZDG_1996,GJN10}:
\begin{equation}
\label{G}
    G
    =
    P_{11} + P_{12}K(I-P_{22}K)^{-1}P_{21}
	=:
    {\rm LFT}(P,K).
\end{equation}
For the purposes of Section~\ref{sec:QYK}, we will now briefly review the classical Youla-Ku\v{c}era parameterization of stabilizing controllers together with related notions. The latter include stabilizability, detectability, internal stability, coprime factorizations and matrix fractional descriptions (MFDs). Despite the quantum control context, these notions will be used according to their standard definitions in classical linear control theory \cite{ZDG_1996,vid2011}.
\subsection{Stabilizability of Feedback Connections}
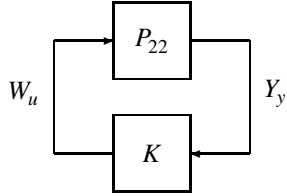
\begin{figure}[htpb]
\begin{center}
\unitlength=1mm
\linethickness{0.6pt}
\begin{picture}(30.00,30.00)
    \put(10,25){\framebox(10,10)[cc]{$P_{22}$}}
    \put(10,10){\framebox(10,10)[cc]{$K$}}
    \put(2,30){\vector(1,0){8}}
    \put(20,30){\line(1,0){8}}
    \put(28,30){\line(0,-1){15}}
    \put(28,15){\vector(-1,0){8}}
    \put(10,15){\line(-1,0){8}}
    \put(2,15){\line(0,1){15}}
    \put(0,23){\makebox(0,0)[rc]{{$W_u$}}}
    \put(30,23){\makebox(0,0)[lc]{{$Y_y$}}}
\end{picture}\vskip-10mm
\caption{Equivalent stabilization diagram.}
\label{fig:system:stab}
\end{center}
\end{figure}

Consider the $(2,2)$ block of the plant transfer matrix $P$ in (\ref{equ:modplnt}) given by
\begin{equation}
    \label{equ:G22}
    P_{22} 
    =
    {\scriptsize\left[
        \begin{array}{c | c}
            A               & B_2 \\
            \hline
            C_2  & D_{22}
        \end{array}
    \right]}.
\end{equation}\noindent
The following lemma provides a necessary and sufficient condition for the internal stability of the feedback system in Fig.~\ref{fig:system}.
\begin{lem} \label{lem:equi} \cite{ZDG_1996} Suppose $(A,B_2,C_2)$ in (\ref{equ:G22}) is stabilizable and detectable. Then the closed-loop system in Fig.~\ref{fig:system} is internally stable if and only if so is the system in Fig.~\ref{fig:system:stab}.
\end{lem}
\vspace{5mm}
\subsection{Stable Factorization}
Let the transfer function $P_{22}$ in (\ref{equ:G22}) have the following coprime factorizations over $\cR \cH_\infty$:
\begin{equation}
\label{fact}
    P_{22}
    =
    NM^{-1}
    =
    \wh{M}^{-1}\wh{N},
\end{equation}\noindent
where the pairs $(N,M)$ and $(\wh{N}, \wh{M})$ of transfer functions in $\cR \cH_\infty$ specify the right and left factorizations, respectively. Then there exist $U, V, \wh{U}, \wh{V} \in \cR \cH_\infty$ which satisfy the B\'{e}zout identities:
\begin{equation}
        \label{equ:Bezout}
        \wh{V}M-\wh{U}N = I,
        \qquad
        \wh{M}V - \wh{N} U = I.
\end{equation}\noindent
The following lemma provides conditions for the existence of stable coprime factors for the system $P_{22}$.
\begin{lem} \cite{ZDG_1996}
	\label{lem:realization}
    Suppose $(A,B_2,C_2)$ in (\ref{equ:G22}) is stabilizable and detectable. Then the coprime factorizations of $P_{22}$ over $\cR \cH_\infty$, described by (\ref{fact}), (\ref{equ:Bezout}), can be realized by
choosing
    \begin{align}
    	\label{equ:rightcopairs}
        {\scriptsize\begin{bmatrix}
               M & U\\
               N  & V
        \end{bmatrix}}
        & =
        {\scriptsize\left[
        \begin{array}{c|c c}
              A+B_2F & B_2 & -L\\
              \hline
                 F & I & 0\\
             	 C_2+D_{22}F & D_{22} & I
        \end{array}
        \right]},\\
	    \label{equ:leftcopairs}
        {\scriptsize\begin{bmatrix}
            \wh{V} & -\wh{U}\\
            -\wh{N}  & \wh{M}
        \end{bmatrix}}
        & =
        {\scriptsize\left[
        \begin{array}{c|c c}
              A+L C_2 & -(B_2+LD_{22}) & L\\
              \hline
                 F & I & 0\\
                 C_2 & -D_{22} & I
        \end{array}
        \right]},
    \end{align}\noindent
    where $F \in \mR^{\mu \times n}$ and $L \in \mR^{n \times \mu}$ are such that both matrices $A+B_2F$ and $A+LC_2$ are Hurwitz. Furthermore, the systems in (\ref{equ:rightcopairs}), (\ref{equ:leftcopairs}) satisfy the general B\'{e}zout identity 
\begin{equation}
    \label{equ:GBezout}
    \begin{bmatrix}
        \wh{V} & -\wh{U}\\
        -\wh{N}  & \wh{M}
    \end{bmatrix}
    \begin{bmatrix}
               M & U\\
               N  & V
    \end{bmatrix}
    =
    \begin{bmatrix}
        I & 0\\
        0 & I
    \end{bmatrix}.
\end{equation}\noindent
%
\end{lem}
\subsection{The Youla-Ku\v{c}era Parameterization}

The main idea of the Youla-Ku\v{c}era parameterization approach is built on stable factorizations (the representation of the transfer matrix of a plant as a fraction of stable rational matrices). By combining the idea of the stable factorizations of a plant with the concept of coprimeness, necessary and sufficient conditions for internal stabilizability are derived in terms of the general B\'{e}zout identities; see Appendix~\ref{app:GBI}. Solving the B\'{e}zout equations, the parameterization of all stabilizing controllers is obtained.  The following lemma applies results on the Youla-Ku\v{c}era parameterization in the frequency domain to the closed-loop system being considered; for more details, see \cite{ZDG_1996} and the references therein.

\begin{lem}
\label{lem:stab_set} \cite{ZDG_1996} 
 Suppose the block $P_{22}$ of the plant transfer matrix $P$ in (\ref{equ:modplnt}) has the coprime factorizations over $\cRH_\infty$, described by (\ref{fact}). 
Then the set of all controllers which achieve internal stability of the closed-loop system is parameterized either by
\begin{equation}
    \label{equ:ctrl}
    K=(U + M Q_r)(V + N Q_r)^{-1},
\end{equation}
with $Q_r \in \cRH_\infty$ satisfying
\begin{equation}
\label{inf1}
    \det (V + N Q_r)(\infty) \ne 0,
\end{equation}
or by
\begin{equation}
    \label{equ:ctrr}
    K=(\wh{V} + Q_{\ell} \wh{N})^{-1}(\wh{U} + Q_{\ell} \wh{M}),
\end{equation}
with $Q_{\ell} \in \cRH_\infty$ satisfying
$
    \det (\wh{V}+Q_{\ell} \wh{N})(\infty) \ne 0
$.
Also, let the auxiliary transfer matrices $U, V, \wh{U}, \wh{V} \in \cRH_\infty$ in (\ref{equ:Bezout}) be  chosen so that 
$    UV^{-1}=\wh{V}^{-1}\wh{U}
$, 
which is equivalent to (\ref{equ:GBezout}); see Appendix~\ref{app:GBI} for more details. Then the set of all stabilizing controllers is parameterized by
\begin{align}
    \nonumber
    K &=(U + MQ)(V + NQ)^{-1}\\\label{equ:stab_ctrl:lft}
     &= (\wh{V} + Q \wh{N})^{-1}(\wh{U} + Q \wh{M})
     = {\rm LFT}(O_y, Q),
\end{align}\noindent
where the parameter $Q\in \cRH_{\infty}$ of these factorizations satisfies
\begin{equation}
\label{inf3}
    \det (V+ N Q)(\infty)\ne 0,
\end{equation}\vskip-1mm\noindent
and $O_y$ is an auxiliary system given by
\begin{equation}
    \label{equ:O_y}
    O_y:=
    {\scriptsize\begin{bmatrix}
        UV^{-1} & \wh{V}^{-1}\\
        V^{-1}    & -V^{-1}N
    \end{bmatrix}}.
\end{equation}
\end{lem}
In what follows, the class of stabilizing controllers will be parameterized using MFDs. However, they can also be parameterized in the LFT framework due to the relationship between MFD and LFT representations; see Appendix~\ref{app:Appndx_LFT} for more details.
\section{Quantum Version of the Youla-Ku\v{c}era Parameterization}\label{sec:QYK}
We will now employ the results of Sections~\ref{sec:LQHO_PR}--\ref{sec:cntlpara} in order to describe stabilizing coherent quantum controllers in the frequency domain. The following lemma represents the $(J_{\mu},J_{\mu})$-unitary condition in terms of the Youla-Ku\v{c}era parameter $Q$ from (\ref{equ:stab_ctrl:lft}).
\begin{lem}
     \label{JJu:coprime}
     Suppose the controller transfer matrix $K$ is factorized according to (\ref{equ:stab_ctrl:lft}). Then $K$ is $(J_{\mu},J_{\mu})$-unitary if and only if the parameter $Q\in \cRH_{\infty}$ satisfies
     \begin{equation}
         \label{equ:Y1_const}
         \Phi
         +
         Q^{\sim}\Lambda
         +
         \Lambda^{\sim}Q
         +
         Q^{\sim}\Pi Q = 0
     \end{equation}\noindent
     for almost all $s \in \mC$, where
     \begin{align}
     	\label{eq:Phi}
     	\Phi
        &:=
        U^{\sim}J_{\mu}U-V^{\sim}J_{\mu} V,\\
     	\Lambda
        &:=
        M^{\sim}J_{\mu}U-N^{\sim}J_{\mu}V,\\
     	\label{eq:Pi}
     	\Pi
        &:=
        M^{\sim}J_{\mu} M - N^{\sim}J_{\mu} N.
	\end{align}\noindent     	     	
Furthermore, under the condition (\ref{inf3}), the feedthrough matrix $K(\infty)$ is well-defined and inherits the $(J_{\mu}, J_{\mu})$-unitary condition from $K$.
\end{lem}
\begin{pf}
    The $(J_{\mu},J_{\mu})$-unitary condition for the controller
    \begin{equation}
    \label{KJK}
        K^{\sim}(s)J_{\mu}K(s) = J_{\mu},
    \end{equation}\noindent
    which must be satisfied for all $s \in \mC$,  can be represented in terms of the right factorization from (\ref{equ:stab_ctrl:lft}) as
    \begin{equation}
    \label{JJ}
        \big(
            (U \!+\! M Q)
             (V  \!+\! N Q)^{-1}
        \big)^{\sim}
        J_{\mu} 
        (U \!+\! M Q)
        (V \!+\! N  Q)^{-1}
        =
        J_{\mu}.
    \end{equation}\noindent
    Using the properties of system conjugation, (\ref{JJ}) is equivalent to
    \begin{equation*}
        (U + M  Q)^{\sim}
        J_{\mu}
        (U + M Q)
        =
        (V + N  Q)^{\sim}
        J_{\mu}
        (V + N Q).
    \end{equation*}  
    After regrouping terms,  this equality  takes the form
    \begin{align*}
        &U^{\sim}J_{\mu}U
        -
        V^{\sim}J_{\mu} V
        +
        Q^{\sim}(M^{\sim}J_{\mu}U - N^{\sim}J_{\mu}V)
        +
        (U^{\sim}J_{\mu} M-V^{\sim}J_{\mu} N)Q
        +
        Q^{\sim}(M^{\sim}J_{\mu} M-N^{\sim}J_{\mu} N )Q = 0.
    \end{align*}\noindent
     This leads to (\ref{equ:Y1_const}), with $\Phi$, $\Lambda$, $\Pi$ given by (\ref{eq:Phi})--(\ref{eq:Pi}). The fact that  the condition (\ref{inf3}) makes the feedthrough matrix $K(\infty)$ well-defined follows directly from (\ref{equ:stab_ctrl:lft}). The $(J_{\mu},J_{\mu})$-unitarity of $K(\infty)$ is established by taking the limit in (\ref{KJK}) as $s\to \infty$.\hfill$\blacksquare$
\end{pf}
Since the $(J_{\mu},J_{\mu})$-unitary condition (\ref{KJK}) and its  equivalent dual form $ KJ_{\mu}K^{\sim} = J_{\mu} $ (cf. (\ref{equ:JJUnit}) and (\ref{dual})) impose the same constraints on the square transfer matrix $K$, a dual condition to the one described in Lemma~\ref{JJu:coprime} holds for the left factorization of the controller in (\ref{equ:stab_ctrl:lft}). This leads to a dual constraint on $Q$, which corresponds to (\ref{equ:Y1_const}), with $\Phi$, $\Lambda$, $\Pi$ being replaced with their counterparts expressed in terms of $\wh{N}$, $\wh{M}$, $\wh{U}$, $\wh{V}$. 
In Appendix~\ref{app:JJYoulaSS}, we also show how expression (\ref{JJ}) imposes constraints on the state-space realization of the Youla parameter.

\begin{thm}
\label{thm:stab_ctrl}
Suppose the block $P_{22}$ of the plant transfer matrix $P$ in (\ref{equ:modplnt}) has the coprime factorizations over $\cRH_\infty$ described by (\ref{fact}).  Also, let the transfer matrices $U,V,\wh{U},\wh{V} \in \cRH_\infty$ in (\ref{equ:Bezout}) satisfy the general B{\'e}zout identity  (\ref{equ:GBezout}). 
 Then the set of all stabilizing $(J_{\mu},J_{\mu})$-unitary controllers $K$ with a well-defined feedthrough matrix $K(\infty)$ is parameterized by (\ref{equ:stab_ctrl:lft}), where the parameter $Q$ belongs to the set
\begin{equation}
 \label{cQ}
    \cQ:=
    \big\{
        Q\in \cRH_{\infty}\
        {\rm satisfying}\ (\ref{inf3})\ {\rm and}\ (\ref{equ:Y1_const})
    \big\}.
 \end{equation}
\end{thm}

\begin{pf}
This theorem is proved by combining Lemmas~\ref{lem:stab_set} and~\ref{JJu:coprime}. Indeed, since the underlying coprime factorizations are assumed to satisfy the general B\'{e}zout identity (\ref{equ:GBezout}), then (\ref{equ:Y1_const})  can be applied to the common parameter $Q$ in (\ref{equ:stab_ctrl:lft}) in order to describe all stabilizing $(J_{\mu},J_{\mu})$-unitary controllers $K$. Their feedthrough matrices $K(\infty)$ are well-defined provided the additional condition (\ref{inf3}) is also satisfied. The resulting class of admissible $Q$ is given by (\ref{cQ}). \hfill$\blacksquare$
\end{pf}

Theorem~\ref{thm:stab_ctrl} provides a frequency domain parameterization of all stabilizing $(J_{\mu},J_{\mu})$-unitary controllers with a well-defined feedthrough matrix and leads to the following theorem.
\begin{thm}
\label{thm:stab_ctrl:coherent}
Under the assumptions of Theorem~\ref{thm:stab_ctrl}, the MFDs (\ref{equ:stab_ctrl:lft}) describe the set of all stabilizing PR quantum controllers $K$, where the parameter $Q$ belongs to the following class $\wh{\cQ}$ defined in terms of (\ref{cQ}):
\begin{align}
	\label{cQhat}
    \wh{\cQ}
    :=
    \big\{
        Q\in \cQ:\          
        \ K(\infty)\in \O(\mu)
    \big\}.
 \end{align}
 \end{thm}
\begin{pf}
The assertion of the theorem is established by combining Theorem~\ref{thm:stab_ctrl} with the frequency domain criterion of PR provided by Lemma~\ref{lem:PR_Freq}.
\end{pf}
The result of Theorem~\ref{thm:stab_ctrl:coherent} provides the parameterization of all stabilizing linear coherent quantum controllers in the frequency domain.

In view of the results of Lemma~\ref{JJu:coprime}, the constraint (\ref{equ:Y1_const}) on the Youla-Ku\v{c}era parameter $Q$ inherits its quadratic nature from (\ref{KJK}). Equation (\ref{equ:Y1_const}) becomes affine (over the field of reals) with respect to $Q$ in a particular case when $\Pi=0$. In view of (\ref{fact}), the transfer function $\Pi$ in (\ref{eq:Pi}) is representable as
$
    \Pi = M^{\sim}(J_{\mu} - P_{22}^{\sim}J_{\mu} P_{22})M
$, 
and hence, it vanishes if the block $P_{22}$ of the plant is $(J_{\mu}, J_{\mu})$-unitary. The following example shows that there exists a PR plant $P$ such that its block $P_{22}$ is $(J_{\mu}, J_{\mu})$-unitary.
\begin{exmp}
Suppose $P$, defined in (\ref{equ:modplnt}), represents a one-mode LQSS with the associated $(D,M,R)$ parameterization (\ref{equ:ABCD}) and dimension $m=4$. Also, assume that the corresponding matrix $D= \diag(D_{11},D_{22})$ and $B_1 J_2 B_1^\rT=0$ ($\det B_1 =0$). It can be shown by inspection that the block $P_{22}$, given in (\ref{equ:G22}), also represents a LQSS. 
\end{exmp}\noindent
However, in view of the results of the following lemma, Corollary~\ref{cor:unstab} shows that the corresponding feedback connection in Fig.~\ref{fig:system} cannot be stabilized by a linear coherent quantum controller.
\begin{lem}
	\label{lem:JJUS}
	Suppose the block $P_{22}$ of the plant transfer matrix $P$ in (\ref{equ:modplnt}) is a $(J_{\mu}, J_{\mu})$-unitary system. Then there exists no $(J_{\mu}, J_{\mu})$-unitary controller $K$ that can stabilize the system in Fig.~\ref{fig:system:stab}.
\end{lem}
\begin{pf}
The system matrix $\bA$ for the closed-loop system in Fig.~\ref{fig:system:stab} can be calculated as
\begin{align}
	\label{eq:app:bA}
	\bA = 
	\left[
	\begin{array}{c c}
	A + B_2 \Delta d C_2
	&
	B_2 \Delta c	
	\\
	b D_{22} \Delta d C_2 + b C_2
	&
	a+b D_{22} \Delta c
	\end{array}
	\right] 
	,
\end{align}
\noindent
where the minimal realizations of the block $P_{22}$ and the controller $K$ are 
$
\scriptsize
P_{22}= 
\left[
	\begin{array}{c | c}
		A & B_2\\ \hline C_2 & D_{22}
	\end{array}
\right] 
$ 
and 
$\scriptsize 
K = 
\left[
	\begin{array}{c | c}
		a & b\\ \hline c & d
	\end{array}
\right] 
$.
Also, $\Delta := (I-dD_{22})^{-1}$. Since $P_{22}$ and $K$ are $(J_{\mu}, J_{\mu})$-unitary, there exist unique non-singular matrices $\Theta_1 \in \mA_{n_1}$ and $\Theta_2 \in \mA_{n_2}$, where $n_1$ and $n_2$ are the McMillan degree of $P_{22}$ and $K$ \cite{AKhP16} such that:
\begin{align}
	\label{eq:JJS1}
	 C_2 &= D_{22} J_{\mu} B_2^\rT \Theta_1,\quad	
	0 = A^\rT \Theta_1 + \Theta_1 A + C_2^{\rT} J_{\mu} C_2,
	\\
	\label{eq:JJS3}
	 c &= dJ_{\mu} b^\rT \Theta_2,\qquad 
	 0 = a^\rT \Theta_2 + \Theta_2 a + c^{\rT} J_{\mu} c.
\end{align}	 
Then, in view of (\ref{eq:JJS1}) and (\ref{eq:JJS3}), 
\begin{align}
	\nonumber
	\bA^\rT \Theta + \Theta \bA 
	&=
	\left[
	\begin{array}{c c}
	A + B_2 \Delta d C_2
	&
	B_2 \Delta c	
	\\
	b D_{22} \Delta d C_2 + b C_2
	&
	a+b D_{22} \Delta c
	\end{array}
	\right]^\rT
	\Theta
	+
	\Theta
	\left[
	\begin{array}{c c}
	A + B_2 \Delta d C_2
	&
	B_2 \Delta c	
	\\
	b D_{22} \Delta d C_2 + b C_2
	&
	a+b D_{22} \Delta c
	\end{array}
	\right]  \\ 
	\label{equ:temp:JJ}
	&=		
	\left[
	\begin{array}{c c}
	\Theta_1 B_2 \Xi \ B_2^\rT \Theta_1
	&
	\Theta_1 B_2 \Xi D_{22}^\rT b^\rT \Theta_2
	\\
	\Theta_2 b D_{22} \Xi B_2^\rT \Theta_1
	&
	\Theta_2 b D_{22} \Xi D_{22}^\rT b^\rT \Theta_2
	\end{array}
	\right], 
\end{align}
where $\Theta:=\diag (\Theta_1,\Theta_2)$, $\Xi:= J_\mu + \Delta d D_{22} J_\mu + J_\mu D_{22}^\rT d^\rT \Delta^\rT$, and use is made of 
\begin{equation*}
	D_{22}^\rT J_\mu D_{22}=D_{22} J_\mu D_{22}^\rT=J_\mu,\qquad d^\rT J_\mu d=d J_\mu d^\rT=J_\mu. 
\end{equation*}
Also, multiplying by $\Xi$ on the left and on the right by $\Delta^{-1}$ and $\Delta^{-\rT}$, respectively,  we have
\begin{align*}
	\Delta^{-1} \Xi \Delta^{-\rT}
	&= (I-d D_{22})J_\mu (I-d D_{22})^\rT + d D_{22} J_\mu (I-d D_{22})^\rT + (I-d D_{22}) J_\mu D_{22}^\rT d^\rT \\
	&= J_\mu -dD_{22}J_\mu - J_\mu D_{22}^\rT d^{\rT} + J_\mu + d D_{22} J_\mu - J_\mu + J_\mu D_{22}^\rT d^\rT - J_\mu \\
	&=0.
\end{align*}
This implies that $\Xi$ and, consequently, the right-hand side of (\ref{equ:temp:JJ}) is zero. Then there exists a matrix $R \in \mS_{n_1+n_2}$ for which $\bA = \Theta R$. Since $\bA$ is similar to a Hamiltonian matrix, its spectrum is symmetric about the imaginary axis; therefore, the closed-loop system in Fig.~\ref{fig:system:stab} cannot be asymptotically stable. \hfill$\blacksquare$
\end{pf}
\begin{cor}
\label{cor:unstab}
Suppose the block $P_{22}$ of the plant transfer matrix $P$ in (\ref{equ:modplnt}) is a $(J_{\mu}, J_{\mu})$-unitary system. Then there exists no linear coherent quantum controller $K$ that can stabilize the system in Fig.~\ref{fig:system}.
\end{cor}
\begin{pf}
In view of the results of Lemma~\ref{lem:JJUS}, the assertion of the corollary is established by combining Theorem~\ref{thm:stab_ctrl:coherent} with the results of Lemma~\ref{lem:equi}.\hfill$\blacksquare$
\end{pf}
Corollary~\ref{cor:unstab} can be considered as a no-go result for linear quantum system stabilization.
\section[Coherent Quantum Weighted $\mathcal{H}_2$ and $\mathcal{H}_\infty$ Control Problems]{Coherent  Quantum Weighted $\mathcal{H}_2$ and $\mathcal{H}_\infty$ Control Problems in the Frequency Domain}\label{sec:H2}
The following lemma, which is given here for completeness, employs the factorization approach in order to obtain a more convenient representation of the closed-loop transfer function when the controller is represented in terms of the Youla-Ku\v{c}era parameterization.
\begin{lem}
    \label{lem:closed-loop}
    Under the assumptions of Theorem~\ref{thm:stab_ctrl}, for any stabilizing  controller $K$ parameterized by (\ref{equ:stab_ctrl:lft}), the corresponding closed-loop transfer matrix $G$ in (\ref{G}) is representable as
    \begin{equation}
    \label{GTTT}
        G
        =
        T_0 + T_1 Q T_2,
    \end{equation}\noindent
    where
    \begin{align}
    		\label{eq:T_0_eq:T_2}
    		\!\!\!\!\!\!T_0
                \!:=\!
                    P_{11}\! +\! P_{12}U\wh{M}P_{21},\quad
    		T_1
                \!:= \!
                    P_{12} M,\quad
    		T_2
                \!:=\!
                    \wh{M}P_{21}.\!\!\!\!
    	\end{align}
\end{lem}
\begin{pf}
    By substituting $P_{22}$ from (\ref{fact}) and $K$ from (\ref{equ:stab_ctrl:lft}) into (\ref{G}), it follows that
    \begin{align*}
        G = &  P_{11} +  P_{12}(U + MQ)(V + NQ)^{-1} 
         (I-NM^{-1}(U + MQ)(V + NQ)^{-1})^{-1}P_{21}\\
        =& P_{11} + P_{12}(U + MQ)
        (V \!+\! NQ\!-\!NM^{-1}(U \!+\! MQ))^{-1}P_{21}\\
        =& P_{11} + P_{12}(U + MQ)
        (V -NM^{-1}U)^{-1}P_{21}\\
        =& P_{11} + P_{12}(U + MQ)\wh{M}P_{21},
    \end{align*}
    which leads to (\ref{GTTT}) and (\ref{eq:T_0_eq:T_2}).
    Here, use is made of the relations
    $
        V -NM^{-1}U = V - \wh{M}^{-1}\wh{N}U = V - \wh{M}^{-1}(\wh{M}V-I) = \wh{M}^{-1}
    $
    which are obtained from (\ref{fact}) and (\ref{equ:Bezout}). \hfill$\blacksquare$
\end{pf}
Lemma~\ref{lem:closed-loop} allows the following coherent quantum weighted $\mathcal{H}_2$ and $\mathcal{H}_\infty$ control problems to be formulated in the frequency domain.
\subsection{Coherent Quantum Weighted $\cH_2$ Control Problem}
Using the representation (\ref{GTTT}), we formulate a coherent quantum weighted $\cH_2$ control problem as the constrained minimization problem
    \begin{align}
    	\label{H2}
        E:= \|W_{\rm out}G W_{\rm in}\|_2^2
        = \|\mbT_0 + \mbT_1Q\mbT_2\|_2^2
        \longrightarrow
        \min
    \end{align}\noindent
with respect to $Q \in \wh{\cQ}$, where the set $\wh{\cQ}$ is given by (\ref{cQhat}). Here,
\begin{equation}
	\label{eq:bT_0_eq:bT_2}
	\mbT_0
    :=
    W_{\rm out}T_0W_{\rm in},
    \qquad
	\mbT_1
    :=
    W_{\rm out}T_1,
    \qquad
	\mbT_2:=T_2W_{\rm in},
\end{equation}\noindent
where $T_0$, $T_1$, $T_2$ are defined by (\ref{eq:T_0_eq:T_2}). Also, $W_{\rm in}, W_{\rm out}\in \cRH_{\infty}$ are given weighting transfer functions for the closed-loop system $G$ which ensure that $\mbT_0 + \mbT_1Q\mbT_2 \in \cRH_2$.
    The $\cH_2$-norm $\|\cdot \|_2$ is associated with the inner product
$
    \bra \Gamma_1,\Gamma_2 \ket:=\frac{1}{2\pi}\int_{-\infty}^{+\infty} \bra \Gamma_1(i\omega),\Gamma_2(i\omega) \ket_{\rm F} \rd \omega
$, where $\bra \cdot , \cdot \ket_{\rm F}$  refers to the Frobenius inner product of complex matrices.
By using the standard properties of inner products in complex Hilbert spaces \cite{reed1980}, the cost functional $E$ in (\ref{H2}) can be represented as
\begin{equation}
\label{prob:cost:H2}
    E
    =
    \|\mbT_0\|_2^2
    +
    2\Re \bra \wh{\mbT}_0, Q\ket
    +
    \bra
        Q,
        \wh{\mbT}_1 Q \wh{\mbT}_2
    \ket,
\end{equation}\noindent
where
\begin{equation}
\label{T0_T2}
	\wh{\mbT}_0 :=\mbT_1^{\sim}\mbT_0\mbT_2^{\sim},
\qquad
	\wh{\mbT}_1:=\mbT_1^{\sim}\mbT_1,
\qquad
	\wh{\mbT}_2:=\mbT_2\mbT_2^{\sim}.
\end{equation}\noindent
In comparison to the original coherent quantum LQG (CQLQG) control problem \cite{NJP_2009}, the coherent quantum weighted $\mathcal{H}_2$ control problem (\ref{H2}) allows for a frequency domain weighting of the cost.
\subsection{ Coherent Quantum Weighted $\cH_\infty$ Control Problem}
Similarly to (\ref{H2}),  a coherent quantum weighted $\mathcal{H}_\infty$ control problem is formulated as the constrained minimization problem
  \begin{equation}
  \label{Hinf}
        \|G\|_\infty
        =
        \|\mbT_0 + \mbT_1Q\mbT_2\|_\infty
        \longrightarrow
        \min
  \end{equation}\noindent
with respect to $Q\in \wh{\cQ}$, where the set $\wh{\cQ}$ is defined by (\ref{cQhat}). Here, $\mbT_0$, $\mbT_1$ and $\mbT_2$ are given by  (\ref{eq:bT_0_eq:bT_2}),  where,  this time,  the weighting transfer functions   $W_{\rm in}, W_{\rm out} \in \cRH_{\infty}$ are not necessarily strictly proper. Recall that the norm in the Hardy space $\cH_\infty$ is defined by
     $
        \|\Gamma\|_\infty
        :=
        \sup_{\omega \in \mR}
        \sigma_{\max}(\Gamma(i\omega))
     $,
where $\sigma_{\max}(\cdot)$ denotes the largest singular value of a matrix. 
Note that both problems (\ref{H2}) and (\ref{Hinf}) are organised as constrained versions of the model matching problem \cite{francis87}. Since the $\cH_2$ control problem is based on a Hilbert space norm, its solution can be approached by using a variational method in the frequency domain, which employs differentiation of the cost $E$ with respect to the Youla-Ku\v{c}era parameter $Q$ and is qualitatively different from the state-space techniques of \cite{VP_2013a}. 
\section{Projected Gradient Descent Scheme for the Coherent Quantum Weighted $\cH_2$ Control Problem}\label{sec:PGS}

Suppose the set $\wh{\cQ}$ in (\ref{cQhat}) is nonempty, and hence, there exist stabilizing PR quantum controllers for a given quantum plant. Tractable conditions for the existence of such controllers remain an open problem which is not considered here.
By using the representation (\ref{prob:cost:H2}) and regarding the transfer function $Q \in \cRH_{\infty}$ as an independent optimization variable, it follows that the first variation of the cost functional  $E$ in (\ref{H2}) with respect to $Q$ can be computed as
\begin{equation}
\label{Egrad}
\delta E
    =
    \Re
    \bra
        \d E,
        \delta Q
    \ket,
    \qquad
    \d E
    := 2(\wh{\mbT}_0+\wh{\mbT}_1 Q \wh{\mbT}_2),
\end{equation}\noindent
where use is also made of (\ref{T0_T2}). In order to yield a PR quantum controller, $Q$ must satisfy the constraint (\ref{equ:Y1_const}) whose variation leads to
\begin{equation}
    \label{equ:Y2_const:var}
    \delta Q^{\sim}(\Lambda+\Pi Q)+(\Lambda^{\sim}+Q^{\sim}\Pi)\delta Q=0.
\end{equation}\noindent
In view of the uniqueness theorem for analytic functions 
\cite{M65}, the resulting constrained optimization problem can be reduced to that for purely imaginary $s = i\omega$ (with an assumption of analyticity in a strip which includes the imaginary axis for the transfer functions involved), with $\omega\in \mR$. The transfer matrices $\delta Q$, satisfying (\ref{equ:Y2_const:var})  at frequencies  $\omega$ from a given set $\Omega\subset\mR$, form a real subspace of transfer functions
\begin{equation}
\label{cS}
    \!\cS
    \!\!:=\!
    \big\{\!
        X\!\in\!  \cRH_2\!:
        (X^*(\Lambda\!+\!\Pi Q)\!+\!(\Lambda^*\!\!+\!Q^*\Pi)X)\big|_{i\Omega}\!=\!0\!
    \big\}.\!\!\!\!\!\!\!\!\!
\end{equation}\noindent
For practical purposes, the set $\Omega$ is used to ``discretize'' the common frequency range of the given weighting transfer functions $W_{\rm in}$,  $W_{\rm out}$ in the coherent quantum weighted $\cH_2$ control problem (\ref{H2}). A conceptual solution of this problem can be implemented in the form of the following projected gradient descent scheme for finding a critical point of the cost functional $E$ with respect to $Q$ subject to (\ref{equ:Y1_const}) at a finite set of frequencies $\Omega$:
\begin{enumerate}
  \item initialize $Q\in \cRH_{\infty}$ so as to satisfy (\ref{equ:Y1_const}), which yields a stabilizing PR quantum controller; choose the input and output weights $W_{\rm in}(i\omega)$ and $W_{\rm out}(i\omega)$, and assign a discrete frequency array $\omega$ in the set $[-\omega_{\ell},\omega_u]\subseteq \mR$. Set the step size $\alpha >0$.
  \item calculate $\d E(i\omega)$ according to (\ref{Egrad}) for each frequency $\omega \in \Omega$;
  \item \label{step:proj} compute $\delta Q(i\omega) = -\alpha \mathrm{Proj}_{\cS}(\d E(i\omega))$ by using a projection onto the set  $\cS$ and a parameter $\alpha>0$;
  \item update $Q$ to $Q+\delta Q$, and go to the second step.
\end{enumerate}
The gradient projection $\mathrm{Proj}_{\cS}(\d E)$ onto the set $\cS$ in (\ref{cS}) is computed in the third step of the algorithm by solving a convex optimization problem; see Appendix~\ref{app:proj} for more information.
 This algorithm also involves interpolation constraints on transfer functions; see \cite{B90,hel98} for more details. The discrete frequency set $\Omega$ and the step-size parameter $\alpha$ can be chosen adaptively at each iteration of the algorithm. The outcome of the algorithm is considered to be acceptable if $Q$ belongs to the set $\wh{\cQ}$ defined by (\ref{cQhat}) of Theorem~\ref{thm:stab_ctrl:coherent}.
\section{Conclusion}\label{sec:Conclusion}
In view of the PR constraints, the set of all stabilizing linear coherent quantum controllers for a given linear quantum plant has been parameterized using a Youla-Ku\v{c}era factorization approach. This approach has provided a formulation of the coherent quantum weighted $\cH_2$ and $\cH_\infty$ control problems for linear quantum systems in the frequency domain. These problems resemble constrained versions of the classical model matching problem. The proposed framework can also be used to develop tractable conditions for the existence of stabilizing quantum controllers for a given quantum plant, which remains an open problem. This problem is also important for the generation of stabilizing PR quantum controllers as initial approximations for iterative algorithms of CQLQG control design. By developing a numerical algorithm for solving the coherent quantum weighted $\cH_2$ or $\cH_\infty$ control problems, the results of this paper can be used to solve the control problems for real physical systems. A numerical algorithm for solving the weighted $\cH_2$ problem can be designed based on the conceptual scheme proposed in this paper. This is a subject for further research and will be addressed in future. 

\begin{ack}            
AKhS is grateful to Dr. Igor G. Vladimirov for a number of helpful discussions and comments on this work.
\end{ack}
\bibliographystyle{Harvard}        
\bibliography{Biblist}           

\appendix
\renewcommand{\theequation}{\Alph{section}\arabic{equation}}
\setcounter{equation}{0}
\section[Parameterization of Linear Coherent Quantum Feedback Systems]{Parameterization of Linear Coherent Quantum Feedback Systems in Position-Momentum Form}
\label{app:sec:PR_LFT_INV}
The following lemma shows the feedback system in Fig.~\ref{fig:system} represents a LQSS. 
\begin{lem}
	\label{lem:PR_LFT_inv}
	Suppose the LQSSs $P$ and $K$ (with vectors of dynamic variables $X_1$ and $X_2$ satisfying the CCRs (\ref{CCRPK})) are interconnected as shown in Fig.~\ref{fig:system}. Then the resulting closed-loop system is also an LQSS. 
\end{lem}
\noindent
In what follows, we provide a proof for Lemma~\ref{lem:PR_LFT_inv}. In particular, we consider the plant and the controller parameterized by the triples $(D_1,M_1,R_1)$ and $(D_2,M_2,R_2)$. Then we show that the closed-loop feedback system in Fig.~\ref{fig:system} represents a LQSS which can be parameterized by the triple $(D,M,R)$:
\begin{align}
	\label{LFT_D}
	D
	&=
	D_{11} + D_{12} \Delta D_2 D_{21},	
	\\
	\label{LFT_L}
	M 
	&= 
	-\frac{1}{2} B^\rT \Theta^{-1},
	\\
	\label{LFT_R}
	R 
	&=
	\begin{bmatrix}
		R_1 	   & R_{12} \\
		R_{12}^\rT & R_2 
	\end{bmatrix}
	 + \wh{R} ,
\end{align}	
where the matrices $B_1:=2 \Theta_1 M_1^\rT$ and $D_1$ are partitioned as
$B_1=:
	{\scriptsize \begin{bmatrix}
		B_{11} &
		B_{12}
	\end{bmatrix}}$
,
$D_1=:
{\scriptsize
	\begin{bmatrix}
		D_{11} & D_{12}\\
		D_{21} & D_{22}
	\end{bmatrix}}$,
\begin{equation}
	\label{def:Delta}
	\Delta := (I-D_2D_{22})^{-1}, 
	\ \
	B :=
	\begin{bmatrix}
		B_{11} + B_{12} \Delta D_2 D_{21}	
		\\
		B_2 (D_{21} + D_{22} \Delta D_2 D_{21})
	\end{bmatrix},	
	\ \
	B_2:= 2\Theta_2 M_2^\rT.
\end{equation}
Also, the block entries of the symmetric matrix $\wh{R}$ are formulated as
\begin{align*}
	\wh{R}_{11}
	&=	
	\frac{1}{4}
	\Theta_1^{-1} 
	\Big(	 
	B_{12} \Delta D_2 D_{21} J B_{11}^\rT 
	-
	B_{11} J D_{21}^\rT D_2^\rT \Delta^\rT B_{12}^\rT
 	\\
 	&
	\qquad 	\qquad 	
 	+
	B_{12} \Delta D_2 D_{22} J B_{12}^\rT
	-
	B_{12} J D_{22}^\rT D_2^\rT \Delta^\rT B_{12}^\rT
	\Big)
	\Theta_1^{-\rT}	,
	\\
	\wh{R}_{12}&= 
	\frac{1}{4}
	\Theta_2^{-1}	
	B_2
	\Big(			
	D_{22} \Delta D_2 D_{21} J B_{11}^\rT 			
	+
	D_{21} J B_{11}^\rT 
	+
	D_{22} \Delta D_2 D_{22}JB_{12}^\rT	
	\\
 	&
	\qquad 	\qquad 	\quad
	+
	D_{22}JB_{12}^\rT	
	-
	J D_2^\rT \Delta^\rT B_{12}^\rT	
	\Big)
	\Theta_1^{-\rT},
	\\
	\wh{R}_{22} & = 
	\frac{1}{4}
	\Theta_2^{-1} 
	B_2
	\Big(	
	D_{22} \Delta D_2 J 
	-
	JD_2^\rT\Delta^\rT D_{22}^\rT 
	\Big)
	B_2^\rT 
	\Theta_2^{-\rT}.		 
\end{align*}
\begin{pf}
The matrices $A$, $B$, $C$ and $D$ for the closed-loop system can be calculated as
\begin{align}		
	\label{eq:app:ABCD_rep}
	\left[
	\begin{array}{c : c}
		 A & B\\
		 \hdashline
		 C & D\\
	\end{array}	
	\right] 
	\!\!\!=\!\!\!
	\left[
	\begin{array}{c c : c}
	A_1+B_{12} \Delta D_2 C_{21}
	&
	B_{12} \Delta C_2
	&
	B_{11} + B_{12} \Delta D_2 D_{21}	
	\\
	B_2 D_{22} \Delta D_2 C_{21} + B_2 C_{21}
	&
	A_2+B_2 D_{22} \Delta C_2
	&
	B_2 D_{21} + B_2 D_{22} \Delta D_2 D_{21}
	\\ 
	\hdashline
	C_{11}+D_{12} \Delta D_2 C_{12}
	&
	D_{12} \Delta C_2
	&
	D_{11} + D_{12} \Delta D_2 D_{21}
	\end{array}
	\right]
	\!\!,
\end{align}
where the state-space realizations for the plant and the controller
\begin{equation*}
	P
	=
	\left[
	\begin{array}{c : c c}
		 A_1 & B_{11} & B_{12}\\
		 \hdashline
		 C_{11} & D_{11} & D_{12}\\
		 C_{21} & D_{21} & D_{22}\\
	\end{array}	
	\right]
	,\qquad
	K
	=:
	\left[
	\begin{array}{c : c}
		 A_2 & B_2\\
		 \hdashline
		 C_2 & D_2
	\end{array}	
	\right],	
\end{equation*}
are the corresponding parameters for the plant and the controller, which depend on the associated Hamiltonian and coupling parameterizations of these systems, and $\Delta$ is defined in (\ref{def:Delta}). 

We will show that the realization (\ref{eq:app:ABCD_rep}) is associated with a LQSS. 
In the first step, we show that the matrix $D$ is an orthogonal symplectic matrix, that is,
\begin{equation}
	\label{fdth}
	D(I+iJ) D^\rT=I+iJ.
\end{equation}
In (\ref{fdth}) and in what follows the subscripts in the matrices $I_k$ and $J_k$ of order $k$ are omitted for brevity.
The matrices $D_1$ and $D_2$ are the feedthroughs of the LQSSs $P$ and $K$; therefore, it can be shown that
\begin{align}
	\label{eq:semp_K1}
	D_{11} (I+iJ)  D_{11}^\rT + D_{12} (I+iJ)  D_{12}^\rT &= I+iJ, \\ 
	\label{eq:semp_K2}
	D_{11} (I+iJ)  D_{21}^\rT + D_{12} (I+iJ)  D_{22}^\rT &=0,\\     
	\label{eq:semp_K3}	
	D_{21} (I+iJ)  D_{21}^\rT + D_{22} (I+iJ)  D_{22}^\rT &= I+iJ, \\ 
	\label{eq:semp_K4}	
	D_2 (I+iJ)  D_2^\rT &= I+iJ.
\end{align}
In view of (\ref{eq:app:ABCD_rep}), from the left hand side of (\ref{fdth}), it follows that
\begin{align*}	
	D(I+iJ)D^\rT
	&=	
	(D_{11} + D_{12} \Delta D_2 D_{21})(I+iJ) (D_{11} + D_{12} \Delta D_2 D_{21})^\rT \\ 
	&=
	D_{11}(I+iJ) D_{11}^\rT \\
	&\quad 
	+ 
	D_{11} (I+iJ) D_{21}^\rT D_2^\rT \Delta^\rT D_{12}^\rT 	
	+  
	D_{12} \Delta D_2 D_{21} (I+iJ) D_{11}^\rT \\
	&\quad
	+  
	D_{12} \Delta D_2 D_{21} (I+iJ)  D_{21}^\rT D_2^\rT \Delta^\rT D_{12}^\rT\\
	& =
	(I+iJ) 
	-
	D_{12}(I+iJ) D_{12}^\rT 
	\\
	&\quad 
	- 
	D_{12} (I+iJ) D_{22}^\rT D_2^\rT \Delta^\rT D_{12}^\rT 
	- 
	D_{12} \Delta D_2 D_{22} (I+iJ)  D_{12}^\rT\\
	&\quad 
	+  
	D_{12} \Delta D_2 
	\Big(
		(I+iJ) - D_{22} (I+iJ)  D_{22}^\rT
	\Big) 
	D_2^\rT \Delta^\rT D_{12}^\rT\\
	&=
	I+iJ,
\end{align*}
where use is also made of the following equation
\begin{align}	
	\nonumber
	0 
	&= 
	-(I+iJ)  
	-(I+iJ) D_{22}^\rT D_2^\rT \Delta^\rT 
	-\Delta D_2 D_{22} (I+iJ)   \\ \label{equ:keq} 
	&\quad
	+\Delta D_2 \Big((I+iJ) - D_{22} (I+iJ)  D_{22}^\rT \Big) D_2^\rT \Delta^\rT,
\end{align}
\noindent
which, in view of (\ref{eq:semp_K4}), can be proved by multiplying its right-hand side from left and right by $\Delta^{-1}=I-D_2 D_{22}$ and $\Delta^{-\rT}$.

In the second step, we will prove that the non-demolition nature of the output fields of the closed-loop system is preserved. That is,
\begin{equation}
	\label{equ:ndcons}
	C \Theta + D J B^\rT = 0,
\end{equation}
where $\Theta :=\diag(\Theta_1,\Theta_2)$.
The non-demolition conditions for outputs of the plant and the controller imply
\begin{align}
	\label{equ:ndconsP}
	\begin{bmatrix}
		C_{11}\\
		C_{21}
	\end{bmatrix}
	&=
	-
	\begin{bmatrix}
		\big(
			D_{11} J B_{11}^\rT + D_{12} J B_{12}^\rT
		\big)
		\Theta_1^{-1}\\
		\big(
			D_{21} J B_1^\rT + D_{22} J B_2^\rT
		\big)\Theta_1^{-1}
	\end{bmatrix},\\
	\label{equ:ndconsK}
	0 \quad &= \quad C_2\Theta_2 + D_2JB_2^\rT.
\end{align}
Then from the left hand side of the condition (\ref{equ:ndcons})
\begin{small}
\begin{align*}
	\nonumber
	C\Theta+DJB^\rT
	&=
	\begin{bmatrix}
				C_{11} \Theta_1+D_{12} \Delta D_2 C_{21} \Theta_1
				\\
				D_{12} \Delta C_2 \Theta_2
	\end{bmatrix} 	
	+ (D_{11} + D_{12} \Delta D_2 D_{21})J \times
	\\ 
	&\quad
	\begin{bmatrix}
				B_{11}^\rT + D_{21}^\rT D_2^\rT \Delta^\rT B_{12}^\rT
				\\
				\big(D_{21}^\rT + D_{21}^\rT D_2^\rT \Delta^\rT D_{22}^\rT \big) B_2^\rT
	\end{bmatrix}\\
	&=
	\begin{bmatrix}
				-\big( D_{11} J B_{11}^\rT + D_{12} J B_{12}^\rT \big) - D_{12} \Delta D_2 ( D_{21} J B_{11}^\rT + D_{22} J 
				B_{12}^\rT)
				\\
				-D_{12} \Delta D_2 J B_{12}^\rT
	\end{bmatrix}\\ 
	&\quad
	+ (D_{11} + D_{12} \Delta D_2 D_{21})J
	\begin{bmatrix}
				B_{11}^\rT + D_{21}^\rT D_2^\rT \Delta^\rT B_{12}^\rT
				\\
				\big( D_{21}^\rT + D_{21}^\rT D_2^\rT \Delta^\rT D_{22}^\rT \big) B_2^\rT
	\end{bmatrix}\\
	&=
	D_{12}
	\big( 
		-J
		-\Delta d D_{22}J 
		- 
		J D_{22}^\rT d^\rT\Delta^\rT 
		+ 
		\Delta D_2 
		\big(
			J- D_{22} J D_{22}^\rT 
		\big)
		D_2^\rT 
		\Delta^\rT 
	\big)
	\begin{bmatrix}				
		B_{12}^\rT
		\\
		D_{22}^\rT B_2^\rT
	\end{bmatrix}\\
	&=0,
\end{align*}
\end{small}	
\noindent
where use is made of (\ref{eq:semp_K2}), (\ref{eq:semp_K3}), the real part of (\ref{equ:keq}), (\ref{equ:ndconsP}) and (\ref{equ:ndconsK}).

In the final step, in view of (\ref{A_B}), we will compute the Hamiltonian parameter $R$ of the closed-loop system from
\begin{equation}
	\label{eq:app:R}
	\wh{R} =  \frac{1}{2} \Big( \Theta^{-1} A - \frac{1}{2} \Theta^{-1} B J B^\rT \Theta^{-\rT} \Big).
\end{equation}
The realization in (\ref{eq:app:ABCD_rep}) implies
\vspace{-1mm}
\begin{small}
\begin{align*}
	E &:=
	\frac{1}{2} \Theta^{-1} B J B^\rT \Theta^{-\rT}
	\\
	&=  
	\frac{1}{2}
	\begin{bmatrix}
		\Theta_1 & 0      \\
		0 	   & \Theta_2
	\end{bmatrix}^{-1}
	\begin{bmatrix}
		B_{11} + B_{12} \Delta D_2 D_{21}	\\
		B_2 D_{21} + B_2 D_{22} \Delta D_2 D_{21}
	\end{bmatrix}
	J 	
	\begin{bmatrix}
		B_{11} + B_{12} \Delta D_2 D_{21}	\\
		B_2 D_{21} + B_2 D_{22} \Delta D_2 D_{21}
	\end{bmatrix}^\rT
		\begin{bmatrix}
		\Theta_1 & 0      \\
		0 	   & \Theta_2
	\end{bmatrix}^{-\rT}\!\!,	
\end{align*}
\end{small}
where the block entries of the matrix $E$ can be calculated as
\vspace{-1mm}
\begin{align}	
	\label{cL1}
	E_{11} &= \frac{1}{2}
	\Theta_1^{-1}
	\big(
	B_{11}JB_{11}^\rT 
	+ 
	B_{11} J D_{21}^\rT D_2^\rT \Delta^\rT B_{12}^\rT
	+
	B_{12} \Delta D_2 D_{21} J B_{11}^\rT		
	+
	B_{12} \Delta D_2 D_{21} J D_{21}^\rT D_2^\rT \Delta^\rT B_{12}^\rT
	\big)
	\Theta_1^{-\rT},
	\\
	\label{cL2}
	E_{12} 
	&= \frac{1}{2}
	\Theta_1^{-1}
	\big(
	B_{11} J D_{21}^\rT 
	+ 
	B_1 J D_{21}^\rT D_2^\rT \Delta ^\rT D_{22}^\rT 
	+ B_{12} \Delta D_2 D_{21} J D_{21}^\rT  	
	+ B_{12} \Delta D_2 D_{21} J D_{21}^\rT D_2^\rT \Delta ^\rT D_{22}^\rT 			
	\big)
	B_2^\rT
	\Theta_2^{-\rT},
	\\ \label{cL3}
	E_{22} &= \frac{1}{2}
	\Theta_2^{-1}
	B_2
	\big(
	D_{21}JD_{21}^\rT
	+ 
	D_{21} J D_{21}^\rT D_2^\rT \Delta^\rT D_{22}^\rT 
	+
	D_{22} \Delta D_2 D_{21} J D_{21}^\rT 
	+
	D_{22} \Delta D_2 D_{21} J D_{21}^\rT D_2^\rT \Delta^\rT D_{22}^\rT 
	\big)
	B_2^\rT
	\Theta_2^{-\rT}	
\end{align}
and $E_{12}=-E_{21}^\rT$.
It follows from (\ref{eq:app:R}) that the closed-loop Hamiltonian parameter is given by
\begin{align}
	\label{wtR}
	R 
	&= 
	\frac{1}{2}
	\begin{bmatrix}
		\Theta_1^{-1}A_1 + \Theta_1^{-1} B_{12} \Delta D_2 C_{21}
		&
		\Theta_1^{-1} B_{12} \Delta C_2 \\
		\Theta_2^{-1} B_2 D_{22} \Delta D_2 C_{21} + B_2 C_{21}
		&
		\Theta_2^{-1} A_2
		+
		\Theta_2^{-1} B_2 D_{22} \Delta C_2
	\end{bmatrix}
	-
	\frac{1}{2}E.
\end{align}
Then the one-one block entry of $R$ can be calculated as
\begin{align*}
	R_{11}
	&=
	R_1
	-
	\frac{1}{4}
	\Theta_1^{-1} 
	\Big(
	-
	B_{11} J B_{11}^\rT 
	- 
	B_{12}JB_{12}^\rT 
	\\
	&\qquad \qquad \qquad
	- 
	2B_{12} \Delta D_2 \big(D_{21} J B_{11}^\rT + D_{22} J B_{12}^\rT\big)
	\\
	&\qquad \qquad \qquad +
	B_{11}JB_{11}^\rT 
	+ 
	B_{11} J D_{21}^\rT D_2^\rT \Delta^\rT B_{12}^\rT
	+
	B_{12} \Delta D_2 D_{21} J B_{12}^\rT
	\\		
	& \qquad \qquad \qquad
	+
	B_{12} \Delta D_2 D_{21} J D_{21}^\rT D_2^\rT \Delta^\rT B_{12}^\rT
	\Big)
	\Theta_1^{-\rT}
	\\
	&=	
	R_1
	-
	\frac{1}{4}
	\Theta_1^{-1} 
	\big(
	- 
	B_{12}JB_{12}^\rT 
	- 
	B_{12} \Delta D_2 D_{21} J B_{11}^\rT 
	-
	2B_{12} \Delta D_2 D_{22} J B_{12}^\rT
	\\
	&\qquad \qquad \qquad 
	+
	B_{12} J D_{21}^\rT D_2^\rT \Delta^\rT B_{12}^\rT	
	+
	B_{12} \Delta D_2 D_{21} J D_{21}^\rT D_2^\rT \Delta^\rT B_{12}^\rT
	\big)
	\Theta_1^{-\rT}
	\\
	&=	
	R_1
	-
	\frac{1}{4}
	\Theta_1^{-1} 
	\Big(
	- 
	B_{12} \Delta D_2 D_{21} J B_{11}^\rT 
	+
	B_{11} J D_{21}^\rT D_2^\rT \Delta^\rT B_{12}^\rT	\\
	&\qquad \qquad \qquad 
	-
	B_{12} \Delta D_2 D_{22} J B_{12}^\rT
	+
	B_{12} J D_{22}^\rT D_2^\rT \Delta^\rT B_{12}^\rT\\	
	&\qquad \qquad \qquad 
	+ 
	B_{12} 
	\big(	
	- 
	J 
	-
	\Delta D_2 D_{22} J 
	-
	J D_{22}^\rT D_2^\rT \Delta^\rT
	\\
	&\qquad \qquad \qquad 
	+
	\Delta D_2 D_{21} J D_{21}^\rT D_2^\rT \Delta^\rT 
	\big)
	B_{12}^\rT
	\Big)
	\Theta_1^{-\rT}\\
	&=	
	R_1
	+
	\frac{1}{4}
	\Theta_1^{-1} 
	\big(	 
	B_{12} \Delta D_2 D_{21} J B_{11}^\rT 
	-
	B_{11} J D_{21}^\rT D_2^\rT \Delta^\rT B_{12}^\rT	\\
	&\qquad \qquad \qquad 
	+
	B_{12} \Delta D_2 D_{22} J B_{12}^\rT
	-
	B_{12} J D_{22}^\rT D_2^\rT \Delta^\rT B_{12}^\rT
	)
	\Theta_1^{-\rT},
\end{align*}
\noindent
where use is made of the dependence of $A_k$ on the parameters $(D_k,M_k,R_k)$ for $k=1,2$, (\ref{equ:keq}), (\ref{equ:ndconsP})  and the fact that $\Theta$ is a skew-symmetric matrix. Similarly, it can be shown that the two-two block of $R$ is
\vspace{-3mm}
\begin{align*}
	R_{22}
	&=
	R_2
	-
	\frac{1}{4}
	\Theta_2^{-1} 
	B_2
	\big(
	-	
	J
	- 
	2D_{22} \Delta D_2 J 
	+
	D_{21}JD_{21}^\rT
	+ 
	D_{21} J D_{21}^\rT D_2^\rT \Delta^\rT D_{22}^\rT 
		\\		\nonumber
	& \qquad \qquad \qquad \quad
	+
	D_{22} \Delta D_2 D_{21} J D_{21}^\rT 	
	+
	D_{22} \Delta D_2 D_{21} J D_{21}^\rT D_2^\rT \Delta^\rT D_{22}^\rT 
	\big)
	B_2^\rT 
	\Theta_2^{-\rT}		\\
	&=
	R_2
	+
	\frac{1}{4}
	\Theta_2^{-1} 
	B_2
	\big(	
	D_{22} \Delta D_2 J 
	-
	J D_2^\rT \Delta^\rT D_{22}^\rT 
	\big)
	B_2^\rT 
	\Theta_2^{-\rT}.
\end{align*}
The off-diagonal terms of the closed-loop Hamiltonian parameter are computed as
\begin{align*}
	R_{12}
	&=
	\frac{1}{4}\Theta_1^{-1}
	\big(
	2B_{12} \Delta D_2 J
	-
	B_{11} J D_{21}^\rT
	-
	B_{11} J D_{21}^\rT D_2^\rT \Delta^\rT D_{22}^\rT 
	-
	B_{12} \Delta D_2 D_{21} J D_{21}^\rT 
	-
	B_2 \Delta D_2 D_{21} J D_{21}^\rT D_2^\rT \Delta^\rT D_{22}^\rT 
	\big)
	B_2^\rT
	\Theta_2^{-\rT},
	\\
	&=
	\frac{1}{4}\Theta_1^{-1}
	\big(
	B_{12} \Delta D_2 J
	-
	B_{11} J D_{21}^\rT
	-
	B_{11} J D_{21}^\rT D_2^\rT \Delta^\rT D_{22}^\rT 
	+
	B_{12} \Delta D_2 
	\big( 
	J
	-
	D_{21} J D_{21}^\rT 
	-
	D_{21} J D_{21}^\rT D_2^\rT \Delta^\rT D_{22}^\rT 
	\big)
	\big)
	B_2^\rT
	\Theta_2^{-\rT},
	\\
	&=
	\frac{1}{4}
	\Theta_1^{-1}
	\big(
		B_{12} \Delta D_2 J
		-
		B_{11} J D_{21}^\rT
		-
		B_{11} J D_{21}^\rT D_2^\rT \Delta^\rT D_{22}^\rT 		
		+
		B_{12} 
		\big(
		\Delta D_2 D_{22} J 
		-
		\Delta D_2 
			\big(
					J - D_{22} J D_{22}^\rT
			\big) 
				D_2^\rT \Delta^\rT 
		\big)
		D_{22}^\rT
	\big) 
	B_2^\rT
	\Theta_2^{-\rT},\\
	&=
	\frac{1}{4}
	\Theta_1^{-1}
	\big(
	B_{12} \Delta D_2 J
	-
	B_{11} J D_{21}^\rT
	-
	B_{11} J D_{21}^\rT D_2^\rT \Delta^\rT D_{22}^\rT 
	-
	B_{12} J D_{22}^\rT 
	-
	B_{12} J D_{22}^\rT D_2 \Delta^\rT D_{22}^\rT 
	\big)
	B_2^\rT
	\Theta_2^{-\rT},
\end{align*}
\begin{align*}
	R_{21}
	&=
	\frac{1}{4}
	\Theta_2^{-1}	
	B_2
	\big(			
	D_{22} \Delta D_2 D_{21} J B_{11}^\rT 			
	+
	2 D_{22} \Delta D_2 D_{22} J B_{12}^\rT\\
	&\qquad \qquad \quad 
	+
	D_{21} J B_{11}^\rT  
	+
	2 D_{22} J B_{12}^\rT	
	-
	D_{21} J D_{21}^\rT D_2^\rT \Delta^\rT B_{12}^\rT\\
	&\qquad \qquad \quad  
	-
	D_{22} \Delta D_2 D_{21} J D_{21}^\rT D_2^\rT \Delta^\rT B_{12}^\rT	
	\big)
	\Theta_1^{-\rT}\\
	&=
	\frac{1}{4}
	\Theta_2^{-1}	
	B_2
	\big(			
	D_{22} \Delta D_2 D_{21}J B_{11}^\rT 			
	+
	D_{21} J B_{11}^\rT 
	\\
	&\qquad \qquad \quad 
	+
	2D_{22} \Delta D_2 D_{22} J B_{12}^\rT	
	+
	2D_{22} J  B_{12}^\rT	
	-
	D_{21} J D_{21}^\rT D_2^\rT \Delta^\rT B_{12}^\rT	
	\\
	&\qquad \qquad \quad 
	-
	D_{22} 
				\big( 
					J + JD_{22}^\rT D_2^\rT \Delta^\rT + \Delta D_2 D_{22} J
				\big)
				B_{12}^\rT	
	\big)
	\Theta_1^{-\rT}
	\\
	&=
	\frac{1}{4}
	\Theta_2^{-1}	
	B_2
	\big(			
		D_{22} \Delta D_2 D_{21} J B_{11}^\rT 			
		+
		D_{21} J B_{11}^\rT 
		\\
		&\qquad \qquad \quad  
		+
		D_{22} \Delta D_2 D_{22} J B_{12}^\rT	
		+
		D_{22} J B_{12}^\rT	
		-
		J D_2^\rT \Delta^\rT B_{12}^\rT	
	\big)
	\Theta_1^{-\rT},
\end{align*}
where use is made of (\ref{equ:keq}). It can be shown, by inspection, that $R_{12} = R_{21}^\rT$ and, as a result, the matrix $R$ is symmetric. Then the corresponding $(D,M,R)$ parametrization for the closed-loop system can be formulated as in (\ref{LFT_D})--(\ref{LFT_R}). \hfill$\blacksquare$
\end{pf}
\section{Cholesky-like Factorizations for Skew-Symmetric Matrices} 	\label{App:CLD}
The existence of Cholesky-like factorizations is addressed in the following lemma.
\begin{lem}
	\label{lem:ch_fact}
	Let $\Theta \in \mA_{n}$ be a non-singular matrix. Then there exists a non-singular matrix $\Sigma \in \mR^{n \times n}$ such that $\Theta = \Sigma J_{n} \Sigma^\rT$. 
\end{lem}
\begin{pf}
	As a consequence of spectral decomposition, in the Murnaghan canonical form (see \cite{benner00} and the references therein), there exists a factorization $\Theta = O \Delta O^\rT$, where the matrix $O \in \mR^{n \x n}$ is orthogonal and the matrix $\Delta \in \mR^{n \x n}$ is block diagonal. Each block on the main diagonal of the matrix $\Delta$ has the form $\scriptsize \begin{bmatrix} 0 & \delta_i\\ -\delta_i & 0 \end{bmatrix}$ with $\delta_i >0$, where $\pm i \delta_i$ is a pair of complex conjugate eigenvalues of $\Theta$. Then, there exists a decomposition $\Theta=\Sigma J_{n} \Sigma^\rT$, where the matrix $\Sigma = O {\rm diag}\{ \sqrt{\delta_1}, \sqrt{\delta_1}, \hdots, \sqrt{\delta_n}, \sqrt{\delta_n} \}$ is non-singular.
Also, for any such $\Sigma$, the matrix $\Sigma \wh{\Sigma}^\rT$ leads to the decomposition of $\Theta$, where $\wh{\Sigma} \in \Sp(n,\mR)$. \hfill$\blacksquare$
\end{pf}\noindent 
In view of Lemma~\ref{lem:ch_fact}, any two non-singular matrices $\Theta_1, \Theta_2 \in \mA_{n}$  are related to each other by a non-singular matrix $\wh{\Sigma}$ as $\Theta_1=\wh{\Sigma} \Theta_2 \wh{\Sigma}^\rT$, where $\wh{\Sigma}= \Sigma_1 \Sigma_2^{-1}$ and $\Theta_k=\Sigma_k J_{n} \Sigma_k^\rT$ for $k=1,2$. 
\section{Lemmas on Linear Fractional Transformation}\label{app:Appndx_LFT}
The following lemmas provide relationships between the MFDs and LFT representation of stabilizing controllers in Sections~\ref{sec:cntlpara}  and ~\ref{sec:QYK}.
\begin{lem} \cite{ZDG_1996}
    \label{lem:MFD_LFT}
    Suppose $V$ is invertible. Then the following MFDs are represented as LFTs:
    \begin{align*}
        (U+MQ)(V+NQ)^{-1} &= {\rm LFT}(O_y,Q),\\
        (V+QN)^{-1}(U+QM) &= {\rm LFT}(O_z,Q),
    \end{align*}
    where $O_y$ and $O_z$ are auxiliary systems given by
    \begin{align}
    \label{Oy0}
        O_y &:= \begin{bmatrix}
                    UV^{-1} & M-UV^{-1}N\\
                    V^{-1} & -V^{-1}N
               \end{bmatrix},\\
        \nonumber
               \qquad
\qquad\qquad        O_z & := \begin{bmatrix}
                    V^{-1}U & V^{-1}\\
                    M-NV^{-1}U & -NV^{-1}
               \end{bmatrix}.    \qquad\qquad\ \,
    \end{align}
\end{lem}
The converse of Lemma~\ref{lem:MFD_LFT} also holds under certain conditions on the system $O_y$ which are addressed below.
\begin{lem} \cite{ZDG_1996}
    \label{lem:LFT_MFD}
    Suppose the system $O_y$ is partitioned as $O_y:=\begin{bmatrix} O_{11} & O_{12} \\ O_{21} & O_{22} \end{bmatrix}$. Then the following LFTs are represented as MFDs:
    \begin{description}
        \item[(a)] if $O_{21}$ is invertible, then
            \begin{equation*}
                {\rm LFT}(O_y,Q)=(U+MQ)(V+NQ)^{-1},
            \end{equation*}
            with
            \begin{align*}
                U&=O_{11}O_{21}^{-1},
                \qquad
                M=O_{12}-O_{11}O_{21}^{-1}O_{22},\\
                V&=O_{21}^{-1},
                \qquad\quad\ \
                N=-O_{21}^{-1}O_{22};
            \end{align*}
             \item[(b)] if $O_{12}$ is invertible, then
            \begin{equation*}
                {\rm LFT}(O_y,Q)=(V+QN)^{-1}(U+QM),
            \end{equation*}
            with
            \begin{align*}
                  \qquad U&=O_{12}^{-1}O_{11},
                  \qquad
                  M=O_{21}-O_{22}O_{12}^{-1}O_{11},\\
                  V&=O_{12}^{-1},
                  \qquad\quad\ \
                  N=-O_{22}O_{12}^{-1}.\qquad\quad\ 
            \end{align*}
    \end{description}
\end{lem}
\section{General B\'{e}zout Identity}\label{app:GBI}
For the purposes of Sections~\ref{sec:cntlpara} and \ref{sec:QYK}, the following lemma describes a generalized version of the B\'{e}zout identity.
\begin{lem}
    \label{lem:GBezout}
Suppose $(N,M)$ and $(\wh{N},\wh{M})$ specify the right and left coprime factorizations in (\ref{fact}). Then for any given $U, V, \wh{U}, \wh{V}\in \cR \cH_\infty$, satisfying the B\'{e}zout identities (\ref{equ:Bezout}), there exist their modified versions  $U', V', \wh{U}', \wh{V}' \in \cR \cH_\infty$ which satisfy the general B\'{e}zout identity:
\begin{equation}
    \begin{bmatrix}
        \wh{V}' & -\wh{U}'\\
        -\wh{N}  & \wh{M}
    \end{bmatrix}
    \begin{bmatrix}
               M & U'\\
               N  & V'
    \end{bmatrix}
    =
    \begin{bmatrix}
        I & 0\\
        0 & I
    \end{bmatrix}.
\end{equation}
\end{lem}
\begin{pf}
    Consider the following modifications of $U,V,\wh{U}, \wh{V}$:
    \begin{align}
    \label{new1}
         U' &:= U, \qquad \wh{U}' := \wh{U} - \Ups \wh{M},\\
    \label{new2}
         V' &:= V, \qquad\,\,  \wh{V}' := \wh{V} - \Ups \wh{N},
    \end{align}
    or
    \begin{align}
    \label{new3}
        U' &:= U +  M\Ups, \qquad \wh{U}' := \wh{U},\\
    \label{new4}
        V' &:= V +  N\Ups, \qquad\, \,  \wh{V}' := \wh{V},
    \end{align}
where
    \begin{equation}
    \label{Ups}
        \Ups := \wh{U}V-\wh{V}U.
    \end{equation}
    By using (\ref{equ:Bezout}) and the relation $\wh{M}N = \wh{N}M$ (following from (\ref{fact})), it can be shown that the new factors $U', V', \wh{U}', \wh{V}'$, defined either by (\ref{new1}) and (\ref{new2}) or by (\ref{new3}) and (\ref{new4}) with the same $\Ups$ from (\ref{Ups}),  belong to $\cRH_{\infty}$ and satisfy (\ref{equ:GBezout}). \hfill$\blacksquare$
\end{pf}
Note that the transfer function $\Ups$ in (\ref{Ups}) vanishes if and only if $\wh{U}V=\wh{V}U$ holds, in which case, (\ref{Oy0}) reduces to (\ref{equ:O_y}).
\section{$(J, J)$-Unitary Constraint and Youla Parameter}\label{app:JJYoulaSS}
In what follows, we show how the relation (\ref{JJ}), given in Section~\ref{sec:QYK}, imposes constraints on the state-space realization of the Youla parameter. 
The conditions for the Youla parameter in (\ref{equ:Y1_const}) can be reformulated as
\begin{equation}
	\label{equ:QQs}
	\Bigg(
			\begin{bmatrix}
				U\\
				V
			\end{bmatrix}
			+
			\begin{bmatrix}
				M\\
				N
			\end{bmatrix} Q
	\Bigg)^{\sim}
	J_T
	\Bigg(
			\begin{bmatrix}
				U\\
				V
			\end{bmatrix}
			+
			\begin{bmatrix}
				M\\
				N
			\end{bmatrix} Q
	\Bigg)	
	=0,
\end{equation}		
where $J_T:=\diag(J_\mu,-J_\mu)$. Now, suppose the state-space realization
$ \scriptsize
Q =
		\left[
		\begin{array}{c|c}
			A_Q & B_Q\\
			\hline
			C_Q & D_Q
		\end{array}
		\right]\!\!  \in \cR \cH_\infty,
$
and the conditions of Lemma~\ref{lem:realization} for stabilizability and detectability of $\cP_{22}$ are satisfied.
Here, $A_Q \in \mC^{\kappa \times \kappa}$, and $B_Q$, $C_Q$ and $D_Q$ are complex matrices of appropriate dimensions. Then
\begin{align}
	\label{equ:decomp1}
	0 
	&= 
	\left[
	\begin{array}{c|c}
		A_T & B_T\\
		\hline
		C_T & D_T
	\end{array}
	\right]^{\sim}
	J_T
	\left[
	\begin{array}{c|c}
		A_T & B_T\\
		\hline
		C_T & D_T
	\end{array}
	\right]	
	=
	\left[
		\begin{array}{c c|c}
		A_T 	  &      0       & B_T\\
		-C_T^*J_TC_T &     -A_T^*	 & -C_T^*J_TD_T\\
		\hline
		D_T^*J_TC_T & B_T^* & D_T^*J_T D_T
	\end{array}
	\right].
\end{align}
Here,
{\small
\begin{align}
	A_T 
	&:= 
	\begin{bmatrix}
		A_Q 	& 		0 		& 	0\\
		B_2 C_Q	& 		A		& 	0\\
		0		&		0		&	A
	\end{bmatrix}\!\!,  
	B_T
	:=
	\begin{bmatrix}
		B_Q\\
		B_2D_Q\\
		-L
	\end{bmatrix}\!\!, 
	D_T
	:=
	\begin{bmatrix}
		D_Q\\
		D_{22}D_Q+I
	\end{bmatrix}
	\\
	C_T&:=	
	\begin{bmatrix}
		C_Q & F & F\\
		D_{22}C_Q & C_{22}+D_{22}F & C_{22}+D_{22}F
	\end{bmatrix}
\end{align}
}\noindent
and $\widetilde{A}:=A+B_2F$, and use is made of the realizations of $M, N, U$ and $V$ given in (\ref{equ:rightcopairs}) and standard addition and multiplication operations on the transfer matrices. For Hurwitz matrices $A_Q$ and $\widetilde{A}$, the  block lower triangular matrix $A_T$ is Hurwitz. Consequently, there exists a Hermitian matrix $\Theta_T \in \mC^{(n+n_0) \times (n+n_0)}$ such that
\begin{equation}
	\label{PRAT}
	\Theta_T A_T + A_T^* \Theta_T + C_T^*J_T C_T =0.
\end{equation}
As a result, (\ref{equ:decomp1}) is equivalent to
\begin{align}
\label{equ:diaglz}
	0&=
	\left[
		\begin{array}{c c|c}
		A_T 	  &      0       & B_T\\
		0   	  &     -A_T^*	 & -(\Theta_T B_T + C_T^*J_TD_T)\\
		\hline
		B_T^* \Theta_T + D_T^*J_TC_T & B_T^* & D_T^*J_T D_T
	\end{array}
	\right],
\end{align}
which is derived by applying a similarity transformation 
$\scriptsize
\begin{bmatrix}
	I & 0\\
	-\Theta_T & I
\end{bmatrix}
$
to the transfer function on the right-hand side of (\ref{equ:decomp1}). Then we can apply an additive decomposition on the right-hand side of (\ref{equ:diaglz}) which implies 
\begin{align}
	&0= 
	\left[
		\begin{array}{c|c}
		A_T 	  &    B_T\\
		\hline
		B_T^* \Theta_T + D_T^*J_TC_T & 0
	\end{array}
	\right] +
	\left[
		\begin{array}{c | c}
		A_T 	  &    B_T\\
		\hline
		B_T^* \Theta_T + D_T^*J_TC_T & 0
	\end{array}
	\right]^{\sim}+ D_T^*J_T D_T
	.
\end{align}
Therefore (\ref{equ:diaglz}) is satisfied if 
\begin{align}
	\label{PRDT}
	D_T^*J_TD_T&=0,\\
	\label{PRBT}
	B_T^* \Theta_T + D_T^*J_TC_T&=0.
\end{align}
In fact, for a given matrix $D_Q \in \mC^{\mu \times \mu}$, the  stabilizing problem can be solved by finding a Hurwitz matrix $A_Q \in \mC^{n_0 \times n_0}$, and arbitrary matrices $B_Q \in \mC^{n_0 \times \mu}, C_Q \in \mC^{\mu \times n_0}$ such that the conditions in (\ref{PRAT}), (\ref{PRDT}) and (\ref{PRBT}) are satisfied.
These conditions, except for (\ref{PRDT}), resemble the necessary and sufficient constraints for a minimal state-space realization
$\scriptsize \left[ \begin{array}{c | c} A_T & B_T \\ \hline C_T & D_T \end{array} \right]$ to be PR; see \cite{JNP_2008} for more details.
\section{Computation of $\mathrm{Proj}_\cS(\d_Q  E)$:}
\label{app:proj}
For 
the numerical scheme provided in Section~\ref{sec:PGS}, we define the projection operator as
\begin{equation}
	\label{projopt}
	\mathrm{Proj}_{\cS}(\d_Q  E)
	 := 
	 \argmin \{ \frac{1}{2}|| \d_Q  E-X||_2 \text{\ : \quad } X \in \cS \},
\end{equation}
where $\cS$ is defined in (\ref{cS}).
The solution to this problem is provided in the following lemma. In this lemma the projection operator $\mathrm{Proj}_{\cH_2}:\cL_2 \rightarrow \cH_2$ is defined as
 	\begin{equation}
 		\mathrm{Proj}_{\cH_2} (X) := \cF(\widetilde{x}(t)),
 	\end{equation}
 	where 
 	$
 	\scriptsize 
 	\widetilde{x}(t)
 	:=
 	\left\{ 
 	\begin{array}{c} 
 		x(t) \quad t\>0 \\ 
 		0 \quad \text{o.w.} 
 	\end{array} 
 	\right.
 	$ 
 	is the causal part of the inverse Fourier transform $x(t):=\cF^{-1}(X)$
 	, and $\cF$ denotes the Fourier transform.

\begin{lem}
	\label{lem:proj}
	Consider the projection problem defined in (\ref{projopt}). The solution to this problem can be formulated as
	\begin{align}
		\label{projsol}
		\mathrm{Proj}_{\cS} (\d_Q  E) 
		=  
		\mathrm{Proj}_{\cH_2}  
			P
		,
	\end{align}
 	where $P:=\d_Q E + (\Lambda + \Pi Q) \Upsilon$ and $\Upsilon = \Upsilon^*$ is chosen such that
 	\begin{align}
 		\label{projcons}
	 	\big( \mathrm{Proj}_{\cH_2} P \big )^*
	 	(\Lambda\!+\!\Pi Q)
	 	\!+\!
	 	(\Lambda^*\!+\!Q^*\Pi)
	 	\mathrm{Proj}_{\cH_2} P
	 	=0.
 	\end{align}
\end{lem}
\begin{pf}
In view of the sesquilinear constraint in (\ref{equ:Y2_const:var}), the optimization problem in (\ref{projopt}) can be formulated by applying the Lagrange method as follows:
\begin{equation}
	\label{projoptM}
	 \cE := || \d_Q  E-X||_2^2 - \bra \Upsilon , X^*(\Lambda\!+\!\Pi Q)\!+\!(\Lambda^*\!+\!Q^*\Pi)X \ket	\rightarrow \min,
\end{equation}
over $X \in \cR \cH_2$ and $\Upsilon$, where $\Upsilon^* = \Upsilon$ is the Lagrange multiplier.
The Fr{\' e}chet derivatives of the augmented cost functional $\cE$ defined in (\ref{projoptM}) are given by
	\begin{align}
		\label{optcon1}
		\d_X \cE 
		&= 
		\mathrm{Proj}_{\cH_2}
		\big( 
			X-\d_Q E-(\Lambda
			 + \Pi Q) Y
		\big),\\
		\label{optcon2}
		\d_\Upsilon \cE 
		&= 
		X^*(\Lambda\!+\!\Pi Q)\!+\!(\Lambda^*\!+\!Q^*\Pi)X.
	\end{align}
Since the problem (\ref{projopt}) is a convex optimization problem (this can be shown by decomposing $X$ into real and imaginary parts), the necessary conditions of optimality, that is, $\d_X \cE =0$ and $\d_\Upsilon \cE=0$, coincide with the sufficient conditions of optimality. Then, in view of (\ref{optcon1}) and (\ref{optcon2}), the optimal solution can be given by (\ref{projsol}), where $P$ satisfies the constraint given in (\ref{projcons}). \hfill$\blacksquare$
\end{pf}

Remark: As an alternative, the optimization problem (\ref{projopt}) can also be solved in the time domain. Then, in view of the interpolation constraints for the $(J,J)$-unitary systems, the solution can be transformed to the frequency domain.
\end{document}